**Assessing innovation in the nascent value chains of climate-mitigating technologies**

Zachary H. Thomas[a,b*], Ellen D. Williams[b,c*], Kavita Surana[d,e], Morgan R. Edwards[a,f]

a. Nelson Institute for Sustainability and the Global Environment, University of Wisconsin–Madison, WI, USA
b. Department of Physics, University of Maryland, College Park, MD, USA
c. Earth System Science Interdisciplinary Center, University of Maryland, College Park, MD, USA
d. Institute for Data, Energy and Sustainability, Vienna University of Economics and Business, Vienna, Austria
e. Center for Global Sustainability, School of Public Policy, University of Maryland, College Park, MD, USA
f. La Follette School of Public Affairs, University of Wisconsin–Madison, WI, USA

*Authors contributed equally

Corresponding author: Zachary H. Thomas

Email: zhthomas@wisc.edu

**ABSTRACT**:

Accelerating climate-tech innovation in the formative stage of the technology life cycle is crucial to meeting climate policy goals. During this period, competing technologies are often undergoing major technical improvements within a nascent value chain. We analyze this formative stage for 14 climate-tech sectors using a dataset of 4,172 North American firms receiving 12,929 early-stage private investments between 2006 and 2021. Investments in these firms reveal that commercialization occurs in five distinct product clusters across the value chain. Only 15% of firms develop *end products* (i.e., downstream products bought by consumers), while 59% support these end products through *components*, manufacturing *processes*, or *optimization* products, and 26% develop business *services*. Detailed analysis of the temporal evolution of investments reveals the driving forces behind the technologies that commercialize, such as innovation spillovers, coalescence around a dominant design, and flexible regulatory frameworks. We identify three patterns of innovation: emerging innovation (e.g., agriculture), characterized by recent growth in private investments across most product clusters and spillover from other sectors; ongoing innovation (e.g., energy storage), characterized by multiple waves of investments in evolving products; and maturing innovation (e.g., energy efficiency), characterized by a dominant end product with a significant share of investments in optimization and services.

Understanding the development of nascent value chains can inform policy design to best support scaling of climate-tech by identifying underfunded elements in the value chain and supporting development of a full value chain rather than only end products.

## 1. INTRODUCTION

Many technologies that will be important for climate change mitigation (i.e., climate-tech) are in the early stages of development and deployment and lack established value chains. Public policies can support climate-tech innovation by focusing on these *nascent value chains*. In the U.S., for example, public funding (e.g., through the Infrastructure Investment and Jobs Act, Inflation Reduction Act, and CHIPS and Science Act) offers an opportunity to strengthen climate-tech (1). Research and development funding not only improves the performance of technologies, but also supports the development of the value chains used to deploy them (2). Additionally, innovation across the value chain can strengthen local, employment-generating industries (3–7). However, designing policies to strengthen value chains for innovative, emerging technologies is challenging. As the end product changes with innovation, so does its value chain. Given that half of future net-zero commitments rely on technologies currently in the prototype or demonstration stage (8), and many other technologies are in the formative phase (9,10), policymakers and analysts must specifically focus on developing the nascent value chains around these technologies.

Despite their importance for accelerating innovation, nascent value chains have not been widely studied in academic and policy circles. Previous literature tends to focus on the formative stages of end products or the value chains of more mature technologies. Research on the formative stages of end products (e.g., a wind turbine) has explored how innovation can enable non-incremental change as technologies with a proof-of-concept advance towards performance and scalability (10–14). However, it overlooks how supporting elements (e.g., blades and towers) change alongside these end products. Research on the value chains of more mature technologies focuses on strategies to strengthen established value chains, such as cost reductions from learning-by-doing (3–5). However, it neglects the fact that supporting products such software, hardware, and services along the value chain experience the same innovation barriers, and that the timing of innovations across the value chain can affect diffusion (15–17). As end use and supporting technologies advance from their formative to growth phases, nascent value chains advance to become established value chains.

There are two potential and complementary approaches for examining nascent value chains. The first approach focuses on different product clusters across the value chain. Differences in the composition of these clusters can reveal important features of end use technologies—e.g., the relative role of software and hardware (18), importance of scale and modularity (12,19), and differences in components based on technology complexity (4,12,20). The second approach focuses on the maturity of the value chain. Value chains evolve as technologies move from emerging to maturing innovation. While previous work on value chains largely focuses on patents (17), patterns of investment in climate-tech start-ups can be an indicator of evolving innovation across the value chain (21,22). Start-ups receive capital in various forms, including competitive government grants as well as early-stage private sector equity investments (e.g., angel, venture capital) that can deliver financial returns for investors (21,23–25). Focusing on early-stage private investments can also help to identify gaps that government or private investors can target to strengthen existing value chains.

In this paper, we assess what characterizes nascent value chains for different climate-mitigating technologies and how they evolve and mature as the value chain coalesces to support a particular end product. Specifically, we explore three research questions:

1. What types of product clusters comprise nascent value chains for climate-mitigating technologies, and what are their functional roles?
2. How does investment in different product clusters along the nascent value chain evolve over time and across sectors as climate-mitigating technologies mature?
3. Can we use insights on nascent value chains and their evolution to target public policies to accelerate innovation in climate-mitigating technologies?

To address these questions, we analyze investments in 3,662 North American start-ups (specifically, firms receiving early-stage private investments, such as equity investments in seed, series A, or series B) between 2006 and 2021 (26). To define product clusters, we categorize the short start-up descriptions based on the functional role of the products they develop using a combination of manual and machine-based semantic analysis. We identify five product clusters that are characteristic of each technology sector: business services, optimization, processes, components, and end products. Using data on early-stage private investments, we assess the evolution of product clusters over time within each sector to identify patterns of innovation across the nascent value chain. We find three patterns: maturing, ongoing, and emerging innovation. Policymakers can use this framework to look beyond end products and design policies that will support the development of full value chains.

## 2. METHODOLODY

We used a three-step approach to analyze the evolving relationships among product clusters and funding sources in nascent value chains. First, we developed and analyzed a dataset of climate-tech start-ups in the formative stage of the technology life cycle (Section 3.1). Second, we used qualitative content analysis to categorize start-ups into product clusters (Section 3.2). Third, we evaluated longitudinal patterns of investment across sectors and product clusters (Section 3.3). This approach enabled us to generalize the process by which firms coalesce into a value chain. While the data is proprietary, all code and output data for figures is available in the Supplementary File 1.

### 2.1 Data

We use data from the Cleantech Group i3 database (26) (accessed 19 January 2022) to identify climate-tech firms for our analysis. These proprietary data contain a total of 29,942 firms and 77,570 investments and have been used in multiple research and policy reports (8,25,27). Cleantech Group reports 25 firm characteristics for each firm and 7 investment characteristics for each investment, sourced from a combination of self-reporting by firms and research by Cleantech Group. Notably, the database is focused on firms that are innovative and entrepreneurial and does not report all businesses or entities (such as universities) working in energy and climate innovation.

From the raw database, we identified a subset of firms that were relevant for analyzing early-stage private investments in climate-tech. We exclude late-stage investments (i.e., Series C and subsequent investments), which are associated with the growth stage of the life cycle where firms have proven technologies, products, or business models that are ready for widespread adoption. Firms with missing data on year founded, location, and sector, or without a short description, were not included in our analysis. The final dataset developed for this study includes 12,929 individual investments and 3,662 climate-tech firms, 91% of which are in the U.S. Supplementary Note 1 gives more explanation of our selection and cleaning process to develop this final dataset.

### 2.2 Product clusters

We used qualitative content analysis, an exploratory, inductive approach to interpreting textual data (28,29), to identify product clusters. We started the process with the solar energy sector, for which we initially analyzed the descriptions of 178 firms that received early-stage private investments. (We later extended the time frame of our analysis and augmented this list with an additional 91 firms.) We focused on identifying preliminary patterns in the products they develop, based on their functional characteristics. For example, we clustered firms related to financing, consulting, or installation services. Next, we clustered firms developing software-based techniques for monitoring and optimizing performance. We also clustered firms developing processes (e.g., deposition of solar cell materials), components (e.g., silicon or gallium arsenide cells) to be incorporated in solar power generation systems, and end use consumer products (e.g., mounted solar panels or solar farms).

We then extended our analysis to 13 additional sectors: energy efficiency, smart grids, agriculture, wind, energy storage, transportation, fuel cells and hydrogen, hydro and marine power, geothermal, nuclear, air (e.g., carbon credit services or air filtration products), advanced materials, and other cleantech (which are tangentially related to climate-tech, e.g., computing or robotics). We performed multiple iterations of the inductive process described above as we considered each sector in turn. Within each sector, we first manually reviewed text descriptions of a subset of firms to identify whether the initial clusters were still applicable and if new clusters were needed. Here we ensured that while the clusters were broad (e.g., an end product refers to anything delivered to a consumer), the corresponding text descriptions and keywords were specific to the sector (e.g., gearboxes for wind and fertilizer for agriculture in the components cluster and silicon deposition techniques and converting combustion vehicles to alternative fuels in the processes cluster). We then used custom code developed by our team in R to cluster firms into categories based on keyword patterns relevant for each sector and each product cluster, with manual verification (see Supplementary Note 2).

Our approach resulted in five final product clusters that were adaptable and applicable across technology sectors. First, business services assist with business models and deployment of new products. Second, optimization refers to methods independently applied to physical products to control, monitor, or improve their functions, usually with products in the form of software. Third, processes are techniques used to design or manufacture physical products. Fourth, components are physical products that are assembled into end products. Fifth and finally, end products are the turn-key products that directly lead to emissions reductions and include consumer products as well as energy delivery products. These definitions were developed to identify functional product types as unambiguously yet broadly as possible, with the goal of both representing all elements of the nascent value chain while comparing meaningfully across different technology sectors (see Supplementary Note 3 for examples of products in each cluster).

### 2.3 Early-stage investments

We analyzed two metrics over time: (1) early-stage private investments and (2) the number of firms receiving these funds, aggregating investments per year for each product cluster and sector. We began with a broad analysis across all sectors and then focused on seven sectors with a clear end product and sufficient data for deeper analysis: solar, wind, energy storage, energy efficiency, smart grid, transportation, and agriculture. We present one illustrative example for each pattern of innovation and discuss other sectors in Supplementary Note 5 and descriptive statistics on all sectors in Supplementary Note 1. Three sectors (energy efficiency, energy storage, and smart grid) present evidence of both maturing and ongoing innovation depending on the analysis period and threshold used for defining a decline in investments; for these sectors, we investigate the specific products that constitute each product cluster over time to characterize the pattern of innovation (see Supplementary Note 6). To assess the effects of policies such as the American Recovery and Reinvestment Act (ARRA), and other market factors related to the financial crisis, we examined the difference in investments made between 2006 and 2013, during which 91% of ARRA outlays by the U.S. Department of Energy were spent, and investments made between 2014 and 2021 (30). We combined insights from the literature on each sector with the descriptions of firms and the products they develop to create a more comprehensive picture of how end products evolved over time.

## 3. RESULTS AND DISCUSSION

### 3.1 Investment trends in early-stage climate-tech firms

We find that products throughout the value chain (beyond the end product) represent a significant proportion of innovative activity in climate-tech startups. On average, only 15% of firms develop end products (see Figure 1a). The remaining 85% operate in the other product clusters that represent the value chain (59% develop components, processes, or optimization products; 26% work in business services). The value chain arises as these product clusters coalesce around a dominant end product in each technology sector (e.g., for energy delivery through turbines in the wind sector or in end-use wholesale or retail items such as fuel cells in the fuel cell and hydrogen sector). The temporal variations in early-stage private investment, aggregated across the 14 technology sectors, also reveal different trends in investments by source (see Figure 1b). For example, private investments in business services and optimization increased relative to the other product clusters beginning in 2014. In contrast, the distribution of grant funding across the product clusters changed gradually, even when the absolute level of funding changed markedly (see Supplementary Figure S2c).

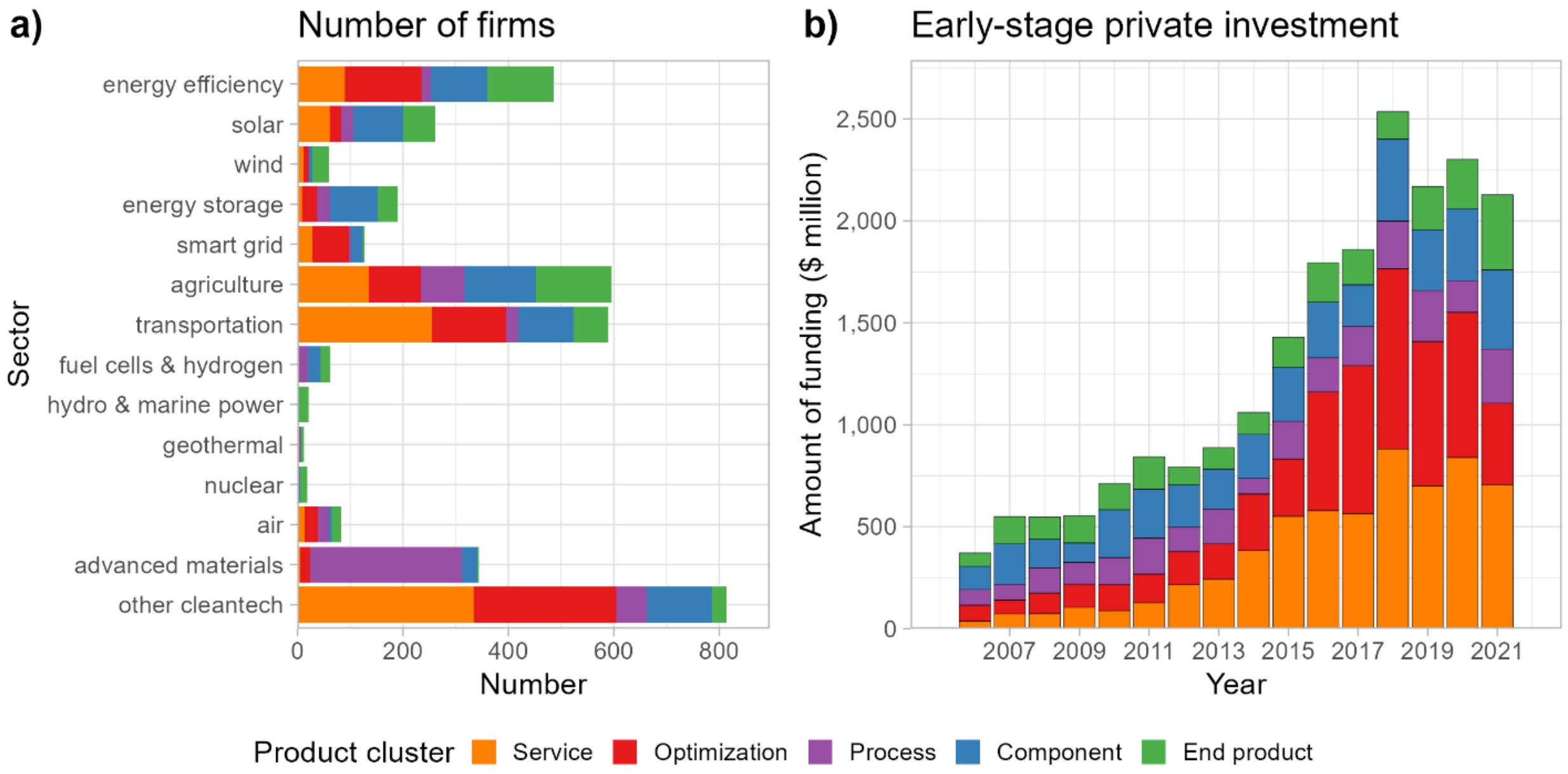

***Figure 1:*** (a) Distribution of the number of climate-tech firms by technology sector and product cluster and (b) temporal trends aggregated across 14 technology sectors for early-stage private investments. Colors indicate the type of product cluster (see Supplementary Table 3 for descriptions).

To investigate trends over time in more detail, we compare investments in each product cluster for the 14 technology sectors across two time periods: 2006–2013 ($5.2 billion total investment) and 2014–2021 ($14.9 billion total investment) (see Figure 1 and Supplementary Note 4). Energy efficiency and solar firms received the most cumulative private investment in the first period (23.9% and 15.9%, respectively). In the second period, these technology sectors were replaced by other cleantech, transportation, and agriculture (34.5%, 27.0%, and 16.8% of cumulative private investment, respectively). Both solar and energy efficiency firms received less funding in the second period, both as a percentage (2.7% and 2.6% of total funding, respectively) and in absolute terms. Private investments in other cleantech and transportation dominated the large increases in total investment in optimization and business services. Similarly, decreasing private investments in solar, energy efficiency, wind, and to a lesser degree energy storage, drove the decrease in total component investments. We show the magnitude of these trends in Supplementary Table 4 and Supplementary Figure S2.

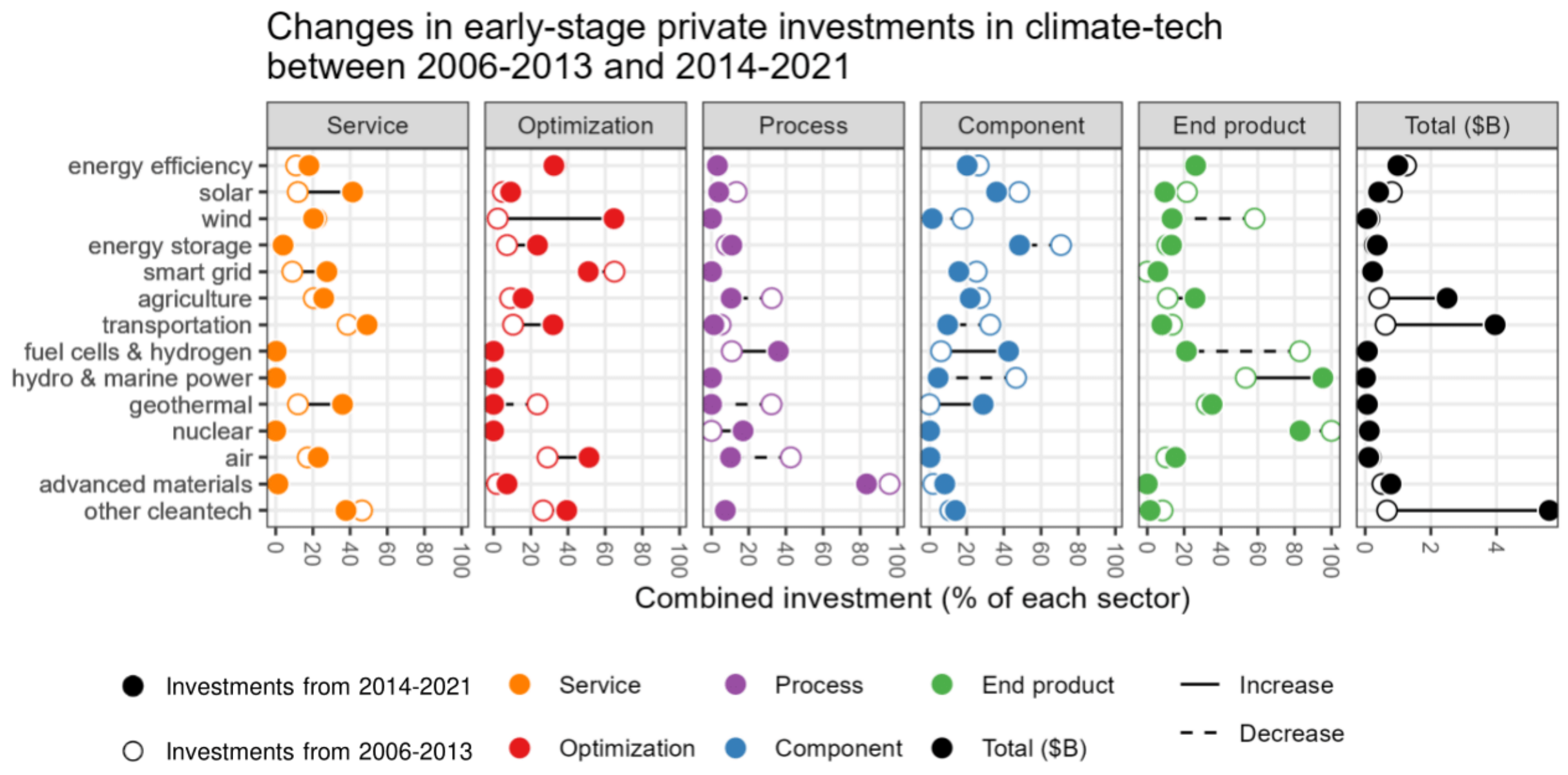

***Figure 2:*** Comparison of cumulative early-stage private investment for the periods 2006–2013 and 2014–2021. Cumulative private investment in the second period (2014–21) was $9.8 billion (188%) larger than the first period (2006–13). This is due to a net effect of a $10.8 billion increase, primarily in other cleantech, transportation, agriculture, and advanced materials, and $1.0 billion decrease, primarily in solar and energy efficiency, with smaller decreases in energy storage, wind, and fuel cells & hydrogen.

### 3.2 Illustrative examples of trends in nascent value chains

Early-stage private sector investment trends reveal distinct patterns of *maturing*, *ongoing*, and *emerging innovation* in nascent value chains across the 14 technology sectors. We categorize sectors into patterns of innovation using the two metrics discussed above (total investment and the number of firms receiving investment) and three indicators: (1) changes in total investment, (2) changes in investment across product clusters, and (3) changes in investment within product clusters. Sectors with maturing innovation experience rapid decreases in total investment while increasing the share of optimization and service products as a single, dominant technology becomes the focus for investors; sectors with ongoing innovation experience multiple waves of investments and gradual evolution of product clusters as technology advances are incorporated; sectors with emerging innovation experience rapid increases total investment and absolute investment in all (or almost all) product clusters. Where the categorization between maturing and ongoing innovation is ambiguous, we also look within product clusters to see whether there is decreasing diversity (a sign of maturing innovation). Patterns of innovation are not necessarily ordinal — for example, a sector can be in a phase of maturing innovation but later receive disruptive investment in new areas and enter a new stage of ongoing innovation.

We discuss three illustrative examples of these innovation patterns: energy efficiency (maturing innovation), energy storage (ongoing innovation), and agriculture (emerging innovation) with details on other sectors in Supplementary Note 5 and Supplementary Figures S3–S9.

### 3.2.1 Maturating innovation: Rise of dominant technologies

The energy efficiency sector encompasses many applications (primarily in building efficiency) and has received large early-stage private investments that began to decline around 2013 (see Figure 3a–b). During the first investment period (2006–2013), the number of firms receiving investment and value of investments increased, following federal legislation that set new efficiency standards (i.e., the 2005 Energy Policy Act and 2007 Energy Independence and Security Act) and directly funded research, development, and deployment of energy efficiency technologies (i.e., through ARRA). Adoption and enforcement of standards was uneven (31), possibly leading to declining support for new energy efficiency technologies after 2013. Later-stage investments, however, remained stable (32). Optimization has been a significant part of the value chain, especially for energy management and control software (EMCS) companies, which make up 77% of optimization companies (see Supplementary Figure S17). This focus may be indicative of continuing improvements in computing power and private investor preference for software companies (2,18).

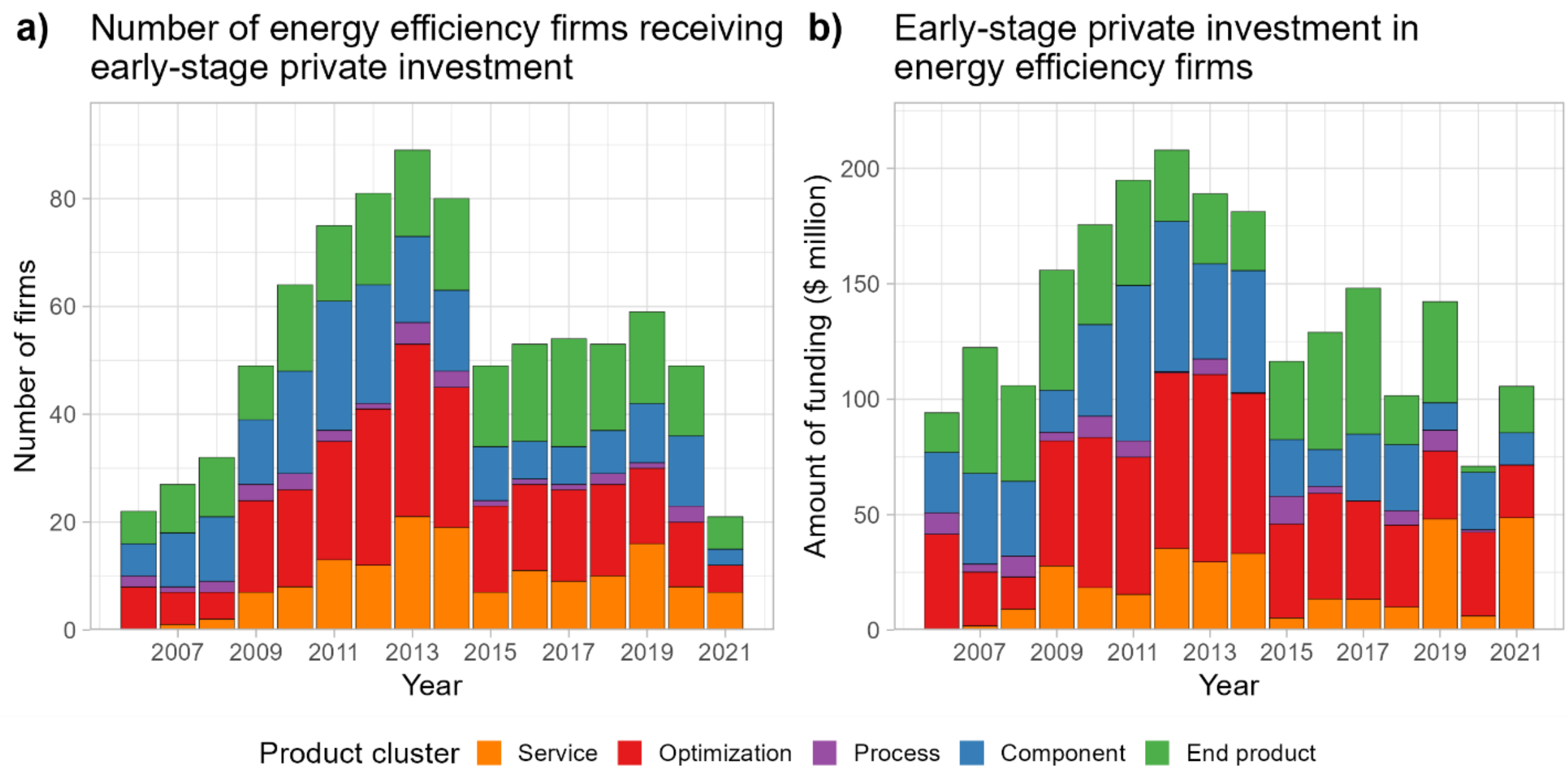

***Figure 3:*** Trends in energy efficiency as an example of maturing innovation in terms of (a) number of firms and (b) total value of early-stage private investments between 2006 and 2021. Colors indicate the type of product cluster within the value chain. For energy efficiency, the cumulative early-stage private investment from 2014–2021 is $200 million (17%) smaller than investment in 2006–2013. Note that vertical axes are scaled differently in the individual charts in Figures 3–5.

Overall, sectors with maturing innovation, including energy efficiency as well as solar and wind, exhibit declines in early-stage private investment and a shift to focus on optimization or service products, coinciding with the establishment of a dominant technology and its associated value chain (see Supplementary Note 5 for discussion of additional examples of maturing innovation). For the energy efficiency sector, this dominant technology reduces energy use in buildings by monitoring and optimizing consumption with digital control technologies (see Supplementary Figure S17 for an overview). Firms across the value chain develop these products, such as component sensors, optimization of EMCS, fully integrated smart apartments as end products, and energy audits as services. Funding in these sectors illustrates the volatile nature of early-stage private-sector investment when a commercially dominant technology is identified and moves into the growth phase and widespread deployment. Without well-designed policies (33), reliance on private investments for alternative technologies may limit innovation or result in premature lock-in of dominant designs.

### 3.2.2 Ongoing innovation: Product cluster stability

Early-stage private sector investment in energy storage, which is important for both electric vehicles and for modernization of the electric power grid, had two periods of increased investment, peaking in 2010 and again 2018 (see Figure 4b). During the first period, there was substantial private sector investment in diverse battery chemistries and flow batteries that was motivated by the potential to compete on cost and performance with Li-ion battery technology (34). However, steady price decreases for Li-ion batteries, particularly in the context of batteries for electric vehicles (35), made competing with Li-ion batteries more challenging for firms developing alternative storage technologies. Funding for these firms was a notable casualty of the investment drop from 2012 to 2014 (see Supplementary Figure S7b). However, from 2017–2018, investment recovered as investors recognized the need for alternative storage technologies for markets such as heavy-duty electric vehicles (36) and utility-scale storage to support renewables integration (37). The initial decrease in funding spurred by a dominant technology was thus premature as it represented only one of the markets (light-duty electric vehicles) in the energy storage sector. To adapt battery technologies to these new markets, innovations such as improvements in active materials and electrolytes led to a 94% increase in investments in process firms before and after 2014 (38).

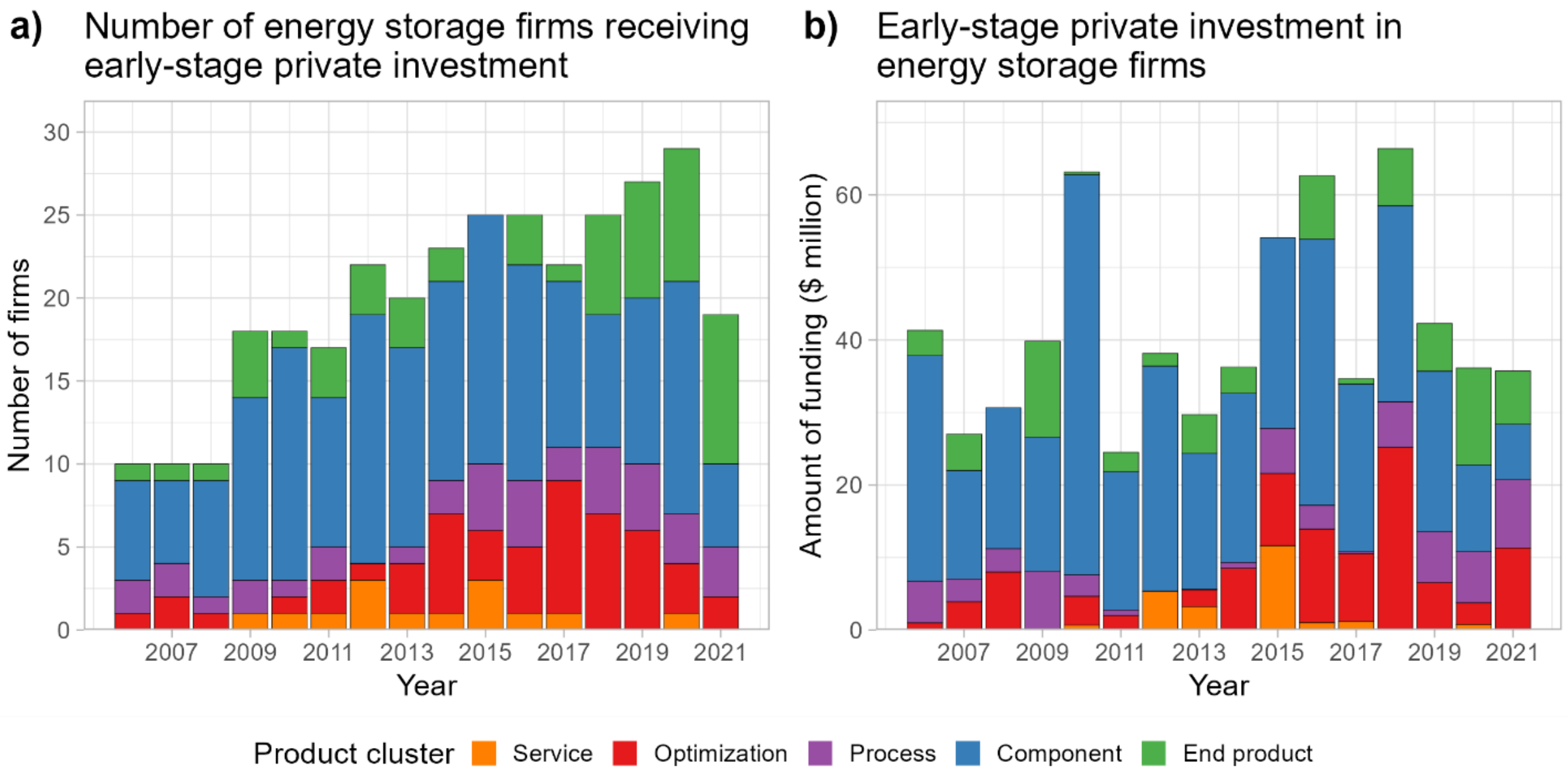

***Figure 4:*** Trends in energy storage as an example of ongoing innovation in terms of (a) number of firms and (b) total value of early-stage private investments between 2006 and 2021. Colors indicate the type of product cluster within the value chain. For energy storage, the cumulative early-stage private investment from 2014–2021 is $6 million (2%) smaller than investment in 2006–2013. Note that vertical axes are scaled differently in the individual charts in Figures 3–5.

An ongoing pattern of innovation can also be seen in other sectors such as the smart grid sector, which exhibits an initial peak in early-stage private investments and number of firms receiving investment in 2011 and smaller peak in 2019 (see Supplementary Figure S7). This contrasts with the maturing pattern of innovation, where sectors experience a continued decline in both the number of firms and amount of early-stage private investment. Additionally, sectors with ongoing innovation experience increases in other metrics that measure innovation, such as the number of investors (e.g., for smart grid, see Supplementary Figure S7b) and the amount of grants (e.g., for energy storage, see Supplementary Figure S6c). Flexible policy frameworks can encourage ongoing innovation by incentivizing firms to develop products for new use cases in the same sector. This is exhibited in the energy storage sector with regional mandates and initiatives to accommodate energy storage in existing energy systems (39,40) and in the smart grid sector with coordination among public agencies and utilities to incentivize adoption of new technologies in distribution networks (41).

### 3.2.3 Emerging innovation: Rapid transformation

Agricultural production is responsible for nearly 10% of all U.S. greenhouse gas emissions (42), but these are primarily direct emissions outside of the energy sector that must be addressed by improvements in agricultural processes. Private sector investment in agricultural climate-tech increased by nearly a factor of five from 2006–2013 to 2014–2021, suggesting a pattern of emerging innovation (see Figures 2 and 5b). All product clusters in the value chain saw significant funding increases, ranging from a factor of two for process and components to a factor of thirteen for end products, primarily from sustainable foods. This rapid evolution coincided with a growing recognition of the potential role for smart agriculture to contribute to climate change mitigation along with other benefits (43–46). Many new products are marketed primarily to save energy or improve yields with ancillary climate benefits, but there are also new efforts to primarily address emissions reductions through innovation in agriculture. (47–49). For example, machine learning to improve farm logistics can reduce time and fuel use during farming operations, with an additional benefit of emissions reductions, while tools that increase the yield of biofuel feedstocks are directly linked to emissions reduction.

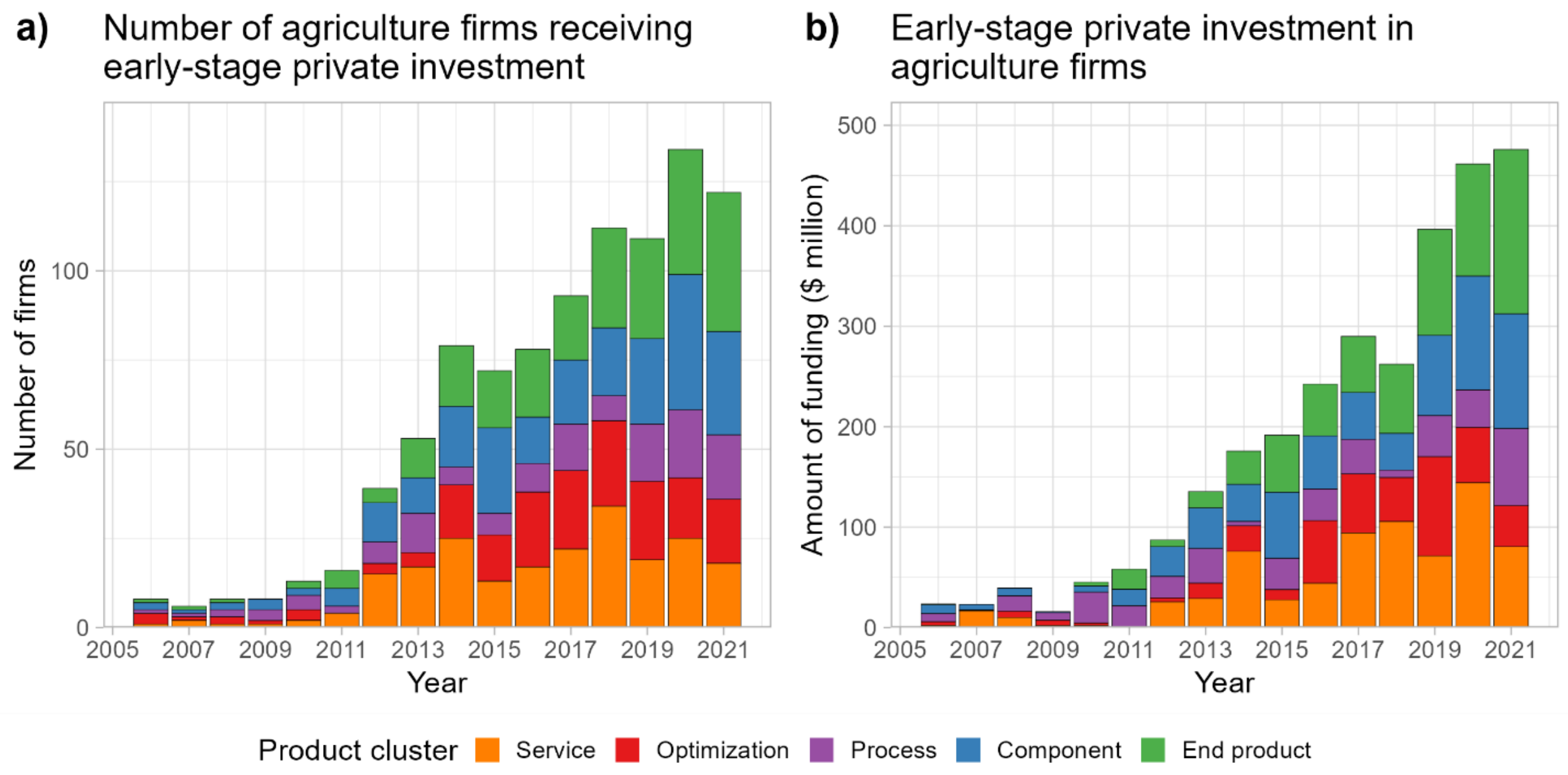

***Figure 5***: Trends in agriculture as an example of emerging innovation in terms of (a) firms and (d) total value of early-stage private investments between 2006 and 2021. Colors indicate the type of product-cluster within the value chain. For agriculture, cumulative early-stage private investment increased by $2.1 billion (495%) from 2006–2013 to 2014–2021. Note that vertical axes are scaled differently in the individual charts in Figures 3–5.

Sectors with emerging innovation can experience rapid transformations in their value chains resulting from spillovers or commercial applications of new technical capabilities (see Supplementary Figure S22). For example, spillovers from machine learning, distributed sensing, and robotics led to innovation in agricultural management tools and precision agriculture, most of which are represented by firms in the services and optimization product clusters. Spillovers from biotechnology enabled development of new bio-agricultural technologies for improved crops and seed and crop treatments (50), which are represented by firms in the components and process product clusters. In transportation, spillovers from machine learning and robotics led to innovations in mobility (e.g., ride-share) and vehicle automation. Transportation and agriculture together represented most of the increase in total investment over our study period. As with agriculture, many of the technologies in the transportation sector that received funding in have a secondary focus on climate change mitigation (see Supplementary Note 6). To ensure continued climate-tech innovation in these sectors, public policies can explicitly incentivize or link climate benefits to the development technologies, including spillover innovations.

## 5. CONCLUSIONS

Large-scale commercialization of technologies to meet climate policy goals will require innovation not only in deliverable end products but also in supporting technologies (2). Public support such as grant funding or targeted demand-pull policies can accelerate this process. However, designing effective policies requires an improved understanding of the nascent value chains in the formative phase of the technology life cycle. In this stage, firms explore new innovations and their applications (13), and both the specific end product(s) and ancillary products of the nascent value chain are in flux. To manage this ambiguity, we used firm-level descriptions to (1) define product clusters consistent across sectors and maturity levels and (2) identify patterns of innovation based on product clusters with common functional characteristics. Our analysis provides examples of how early-stage investment in climate-tech firms, and their specific products, reflects the heterogeneous technology, policy, and market context of each sector. This product- and firm-focused perspective can be a powerful tool for understanding the formative stage of the climate-tech life cycle. To this end, we discuss two key results that inform takeaways for policy to accelerate innovation by supporting nascent value chains.

First, policymakers and analysts need to think beyond end products and explicitly consider the multiple innovative products that comprise nascent value chains. Our analysis contributes to this effort by expanding on prior research that has focused on broad product differentiation such as hardware vs. software (18,51) or on the elements of well-developed value chains (3,4,12,15,16). We identified a novel and consistent set of five functional product clusters across technology sectors — i.e., business services, optimization, process, components, and end products — that receive early-stage private sector investment. These product clusters encompass innovative elements that may combine to deliver a desirable end product, and grow into a well-developed value chain supporting mature operations (6). Granular, technology-specific product cluster data therefore provides unique information on emerging technologies that can help design supportive policy. Policymakers and government agencies that support innovation should develop better tracking and public availability of firm-level data with the granularity to reveal product clusters across the value chain. Public policies can also be more effective when they identify and address gaps in investment in different products within nascent value chains.

Second, policy design needs to be not only technology-specific (10,12,14) but also respond to the different patterns of innovation in the formative stage. Our analysis reveals three different patterns: maturing, ongoing, and emerging innovation. Sectors with maturing innovation experience decreases in early-stage private sector investments with a focus on optimization and services that coincide with a transition to the growth stage, suggesting these sectors would particularly benefit from policy to continue incentivize formative stage innovation, such as grants to new firms. In cases where a dominant technology entering the growth stage represents only one of many possible markets, different policies beyond public investments, such as demand-pull policies, can encourage innovation in ancillary technologies like services and optimization. Sectors with ongoing innovation have resurgent investments across product clusters because of new use-cases for existing technologies. For these cases, policymakers can monitor improvements in technology capabilities to update incentives for innovation. Sectors with emerging innovation are experiencing rapid growth in early-stage private investment, which in the last decade notably resulted from new technology spillovers, with limited policy guidance. Where these investments are driven primarily by non-climate concerns, grants and other policy levers can work to incentivize emissions reductions alongside other innovations.

The formative stage is an uncertain and critical stage in the innovation process, forming a bridge between early research and development and market growth. Encouraging innovation across the nascent value chain during the formative stage can accelerate the process of forming a strong value chain and facilitate widespread deployment of climate-mitigating technologies. The framework we present for analyzing investment trends across the value chain is designed to generalize across different climate-tech sectors and can inform the design of new policies to support a wide range of technologies that deliver climate-mitigating benefits. Future research exploring how the nascent value chain relates to technical, market, and policy influences will further strengthen effective policy design.

## Author contributions

ZHT: Formal analysis (lead), methodology (equal), software (lead), visualization (lead), writing – original draft (equal), writing – review & editing (lead)

EDW: Conceptualization (equal), formal analysis (support), funding acquisition (lead), methodology (equal), project administration (equal), supervision (support), visualization (supporting), writing – original draft (lead)

KS: Conceptualization (equal), formal analysis (support), funding acquisition (supporting), methodology (equal), supervision (support), visualization (support), writing – original draft (support)

MRE: Funding acquisition (support), supervision (equal), writing – review & editing (equal)

## Competing interests

The authors state no competing interests.

## Data availability

The code used to produce the analysis will be available as supplementary information following publication.

# SUPPLEMENTARY INFORMATION

## TABLE OF CONTENTS

## SUPPLEMENTARY NOTE 1: Methods for data cleaning

Cleantech Group's i3 database is one of the most comprehensive sources of information on early-stage investment in climate-tech firms during the formative stage. It categorizes firms into 19 technology sectors (see Supplementary Table 1). Because of its wide usage (1–3), we expect that the i3 database provides the most comprehensive data available on firms working in climate-tech that seek (or receive) early-stage investment. However, the database is based on early-stage climate-tech firms, and thus is does not track all funding for energy and climate innovation (e.g., government grants awarded to recipients that are not these firms) (4).

We considered all firms founded after 1980, located in North America (the region with the largest number of firms in the raw database), and that received early-stage funding (i.e., grant or seed, series A, or series B) between 2006 and 2021 (see Supplementary Table 2). Early-stage investments are generally associated with small firms developing high-risk technologies or business models (5). To ensure that our data represented early-stage investments, we removed data points with large outliers in individual grants and early-stage private investments (see Supplementary Figure S1). We included small investments, including investment amounts recorded as $0. In our manuscript, we present only early-stage private funding (i.e., seed, series A, or series B) and present grants in the Supplementary Information. We make this distinction to discuss grants as a type of policy support that can address gaps in developing nascent value chains. We therefore include more firms in the Supplementary Information, specifically firms that receive only grant funding.

We also undertook additional steps to verify our data and cluster definitions. First, if the text description of the firm was too vague (e.g., listing several different products or no products at all), we supplemented the description with information from the firm's website. Second, if the text description had no clear link to emissions reductions (e.g., grocery delivery services), we removed the firm from our dataset. Finally, if the firm was more closely affiliated with a different technology sector, we updated the sector designation originally provided by Cleantech Group. Additionally, multiple members of our research team verified the qualitative content analysis of product clusters by reviewing each other's categorizations. This collaborative process also led to refinements in cluster definitions and sector-specific keywords.

We also evaluated whether our results were robust to outlier investment rounds and found that removing outliers did not change our results. We define an investment round as the total investment that one company receives on one date, and we define outlier investment rounds to be rounds that are greater than 1.5 times the interquartile range above the 75th percentile investment round for each sector (see Supplementary Figure S1). For example, one solar firm received a $75 million series B investment from one investor in 2008, and another solar firm received six $8.3 million individual series B investments in a single round (totaling $50 million on a single date in 2008). These two rounds are each much larger than the average $2.0 million solar investment round and over 1.5 times the 75th percentile solar investment round of $15.5 million. Excluding the single $75 million series B investment in 2008 and the six $8.3 million series B investments in a single company in 2008 did not change the overall trends of investments in the solar sector. We remove these outlier investment rounds because they are generally characteristic of later stages of innovation.

In total, there were 987 outlier investment rounds, including 3,796 individual outlier investments, corresponding to 20% of the investment data that fit our other selection criteria. Supplementary Table 1 shows the median, mean, and standard deviation on the amount of early-stage private investments in each technology sector, including and excluding outlier investment rounds. While the mean early-stage private investment in each sector is skewed $0.6 million to $17.0 million larger when including outlier investment rounds, trends in investments overall and within different product clusters remain the same (and thus so do sector classifications as emerging, ongoing, or maturing innovation). Supplementary Table 2 provides a summary of the data on firms and investments that were available from Cleantech Group's i3 database and the subset that were included in our analysis. Supplementary Figure S2 shows (a) firms including those that receive grants, (b) amount of early-stage private investments, (c) amount of grants, and (d) number of investors.

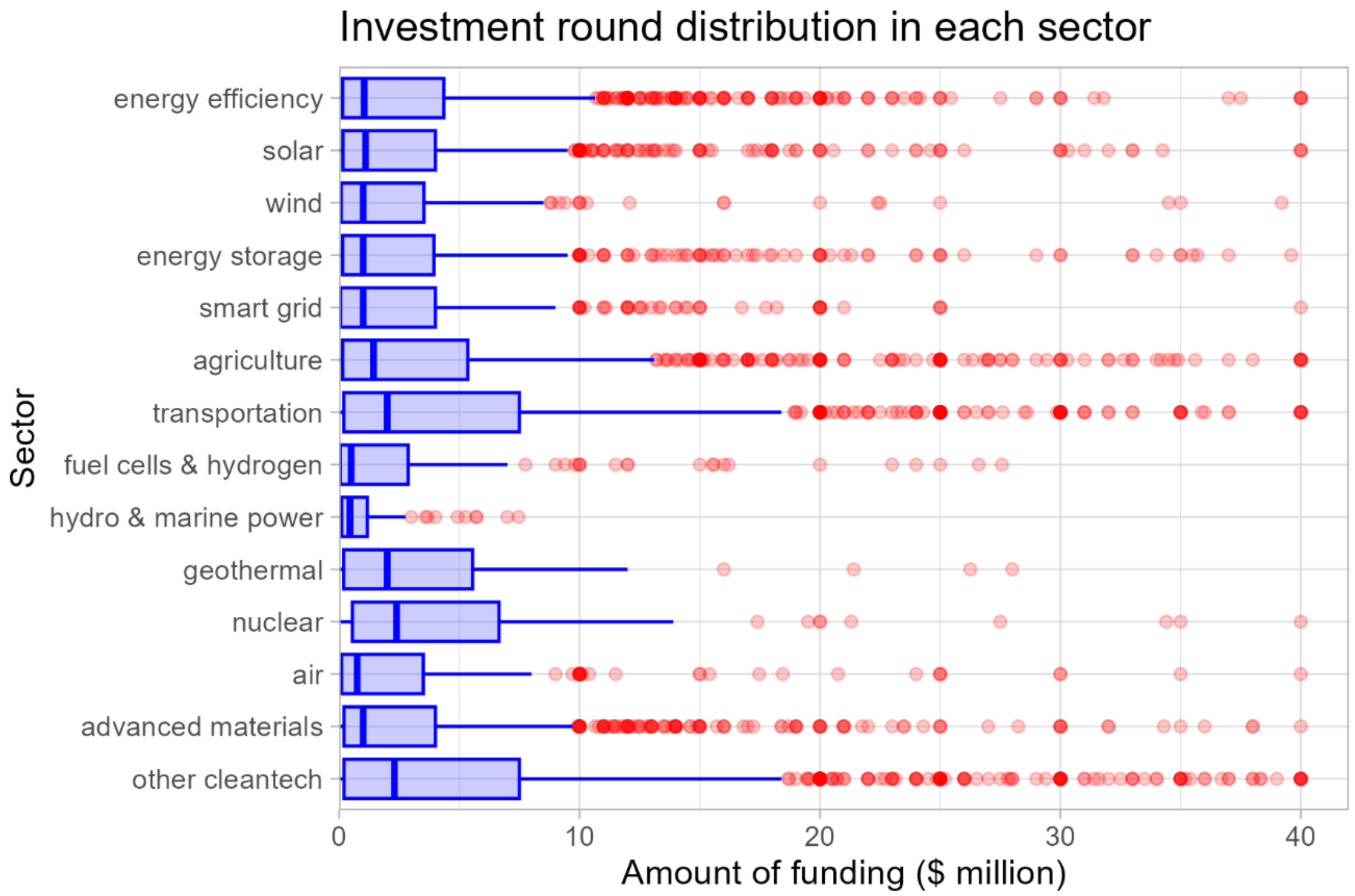

***Figure S1:*** Distribution of grants and early-stage private investment rounds in each sector in Cleantech Group's i3 database. The thick vertical line in each box is the median investment. The difference between the rightmost vertical line (the 75th percentile investment) and the leftmost vertical line in the shaded box is the interquartile range. The horizontal whiskers designate 1.5 times the interquartile range less than and greater than the 25th and 75th percentile investment rounds, respectively. Outlier investment rounds are shown as transparent dots, and opaque dots represent multiple investment rounds of the same amount. The horizontal axis is scaled to $40 million. This omits the largest 193 values, but these values are included in the data in Supplementary Table 1.

***Supplementary Table 1:*** Comparison of median, mean, and variance of grants and early-stage private investments in 14 sectors from the i3 database that we include in our analysis, with and without outliers included. (The five sectors in i3 that we do not analyze are water & wastewater, recycling & waste, conventional fuels, biofuels & biochemicals, and biomass.)

| | *Including outliers* | | | *Excluding outliers* | | |
|---|---|---|---|---|---|---|
| Technology sector | Median ($ million) | Mean ($ million) | Standard Deviation ($ million) | Median ($ million) | Mean ($ million) | Standard Deviation ($ million) |
| Energy efficiency | 1.4 | 2.5 | 3.5 | 0.9 | 1.4 | 1.7 |
| Solar | 1.6 | 3.4 | 5.6 | 0.9 | 1.4 | 1.8 |
| Wind | 1.2 | 2.4 | 3.7 | 0.6 | 1.3 | 1.7 |
| Energy storage | 1.9 | 3.1 | 4.4 | 1 | 1.3 | 1.5 |
| Smart grid | 1.4 | 2.5 | 5.5 | 0.8 | 1.2 | 1.5 |
| Agriculture | 1.2 | 2.8 | 6.1 | 0.7 | 1.3 | 1.7 |
| Transportation | 1.5 | 4 | 20.8 | 1 | 1.8 | 2.4 |
| Fuel cells & hydrogen | 1 | 2.1 | 3.1 | 0.2 | 0.9 | 1.5 |
| Hydro & marine power | 0.7 | 1.3 | 1.8 | 0.4 | 0.7 | 0.7 |
| Geothermal | 2 | 2.6 | 2.7 | 0.7 | 2 | 3 |
| Nuclear | 5.5 | 19.4 | 28.2 | 2 | 2.4 | 3.2 |
| Air | 1.4 | 2.7 | 4 | 0.7 | 1.2 | 1.4 |
| Advanced materials | 1.7 | 2.9 | 5 | 0.9 | 1.4 | 1.6 |
| Other cleantech | 1.7 | 3.6 | 13.5 | 1.2 | 1.9 | 2.2 |

***Supplementary Table 2:*** Summary of firms and investments for each technology sector in the i3 database. The second column lists the number of firms that fit our investment criteria (excluding those that receive outlier investment rounds; see discussion in this Supplementary Note).

| Technology sector | Firms in the i3 database | Firms that meet our investment criteria | Number of grants or early-stage private investments (excluding outliers) | Total amount invested from grants or early-stage private investment (excluding outliers) |
|---|---|---|---|---|
| Energy efficiency | 3,370 | 582 | 1,889 | $2,476 million |
| Solar | 2,463 | 313 | 1,028 | $1,427 million |
| Wind | 872 | 69 | 169 | $219 million |
| Energy storage | 745 | 241 | 771 | $1,017 million |
| Smart grid | 804 | 165 | 468 | $549 million |
| Agriculture | 1,759 | 624 | 2,407 | $3,030 million |
| Transportation | 2,378 | 636 | 2,765 | $4,803 million |
| Fuel cells & hydrogen | 276 | 90 | 215 | $204 million |
| Hydro & marine power | 317 | 35 | 92 | $55 million |
| Geothermal | 130 | 19 | 54 | $104 million |
| Nuclear | 44 | 28 | 93 | $270 million |
| Air | 562 | 117 | 330 | $385 million |
| Advanced materials | 1,602 | 405 | 1,254 | $1,554 million |
| Other cleantech | 1,839 | 848 | 3,246 | $5,883 million |

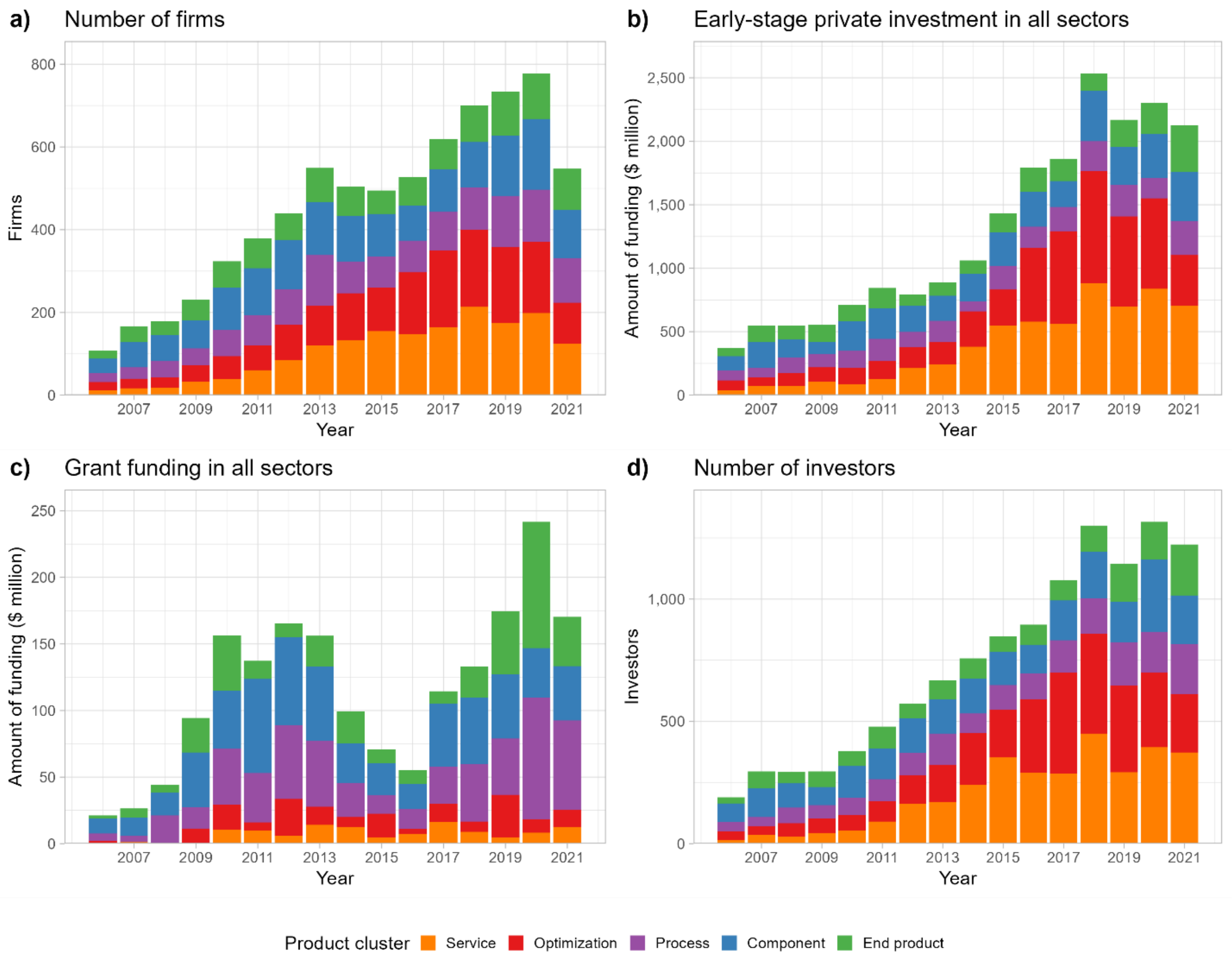

***Figure S2:*** Comparison of number of firms (a) in the seven sectors for which we categorized firms by product cluster, and the early-stage private investments (b), grants (c), and number of investors (d) in these firms between 2006 and 2021. Vertical axes are scaled differently in the individual charts.

## SUPPLEMENTARY NOTE 2: Description of code to categorize firms into product clusters

We used custom code developed by our team (using R version 4.0.2 (6); code is included in the Supplementary Information as Supplementary Data Files) to cluster firms into relevant categories based on keyword patterns relevant for each sector. This allowed us to search for descriptive combinations of key terms rather than isolated words. For example, instead of using a single term such as "silicon", we used expressions such as "residential silicon solar cells" or "silicon deposition" to accurately distinguish different types of products, in this case end-use and processes. To establish a repeatable method for categorizing companies into product clusters, we used this approach to iteratively filter text descriptions for key terms associated with each product.

First, the code in R identified firms that unambiguously belonged to a specific product cluster related to business services, optimization, or processes. Starting with business services, we identified firms with keywords such as "consulting" or "financing." We then identified in the remaining firms (i.e., excluding the business service firms) those with keywords such as "monitoring" or "control" (related to optimization) or with keywords such as "research" or "materials" (related to processes).

Second, we identified in the remaining firms (i.e., excluding firms already identified as business services, optimization, or process) those that developed components and end products. The distinction between components and end products was necessary at the sector level; for example, energy storage systems are end products in the energy storage sector, but storage systems for wind farms are components in the wind sector. To this end, we reviewed the remaining firm text descriptions to identify keywords that describe the end products commonly developed within each sector. In a third step, we continued this iterative process to identify relevant keywords in the remaining firms' descriptions until the uncategorized list of firms had no major repetitive patterns and was small enough to sort by hand (i.e., less than 50).

Finally, after categorizing the 14 sectors into the same five product clusters, our team manually reviewed random subsets of firms in each sector and verified their product cluster categorizations, confirming that the definition of each product cluster could independently describe the products in each sector. Because the categorization of each firm's product cluster is based on a single sentence, we expect that some categorizations do not accurately represent firms that develop multiple products, have changed their product development, or have ambiguously represented their product; however, we expect these examples are infrequent and do not affect overall trends in investments across product clusters and sectors.

## SUPPLEMENTARY NOTE 3: Examples of products in each product cluster

Supplementary Table 3 lists examples of specific product functions in each sector and product cluster. As examples, many of the firms developing end products in the air sector relate to carbon capture and storage (CCS) and several of the firms offering business services in agriculture relate to logistics.

***Supplementary Table 3:*** Selection of frequently occurring key terms in each product cluster and sector. Blank cells indicate that the sector did not have many common products in the corresponding product cluster.

| Sector | End products | Components | Processes | Optimization | Business Services |
|---|---|---|---|---|---|
| Energy efficiency | HVAC systems; waste power generation; building upgrades | LEDs; power electronics; motors; HVAC components | Semiconductor materials; manufacturing process | IoT; metering; smart technologies; monitoring and control for appliances, buildings, lighting, and climate systems | Consulting; audits; incentives to adopt energy efficiency improvements |
| Wind | Turbines; wind farms; portable turbines | Turbine blades, gearboxes | | Weather forecasting; predictive maintenance | Wind farm operators; maintenance |
| Solar | Distributed systems; centralized (PV and CSP) systems | Cells; modules; power electronics; converters, inverters, mounting hardware | Cell production; materials | Site assessment; performance solar trackers | Distribution; installation; financing |
| Energy storage | Storage systems | Batteries (lithium ion, other battery chemistries and flow cells); thermal; capacitors; cathodes anodes; battery cooling | Manufacturing processes; R&D; recycling; battery materials (ceramics, metals) | Smart storage systems; battery system management | Storage asset management |
| Smart grid | Turnkey microgrid systems | Power electronics; sensors | Power cable materials | Sensor networks; controls and management for demand-response, power distribution, distributed energy resources (DER), and integration with energy storage | Wireless networking systems; cloud and edge computing; financial analysis for distributed energy; cybersecurity |
| Transportation | Large vehicles, such as cars, | Powertrains; motors; EV charging | Vehicle conversions; fuel additives | Automation and connectivity; traffic | Mobility, ride sharing, ride |
| | trucks, aircraft; small vehicles such as bicycles, carts | | | management; mapping and tracking | sourcing; fleet management, logistics; parking; delivery; data analytics |
| Agriculture | vertical farm or hydroponics equipment, greenhouses and lighting, field equipment including robotic vehicles | Seeds or seed treatments to increase crop yield; reduce use of fertilizer; pesticides, or herbicides; soil treatment; animal feed; aquaculture; preservatives | Chemical and biochemical processes, products from waste, genomics | Precision agriculture; field robotics; farm processes such as irrigation | Farm management; market connectivity; product transport software (omitted: retail food deliveries) |
| Fuel cells & hydrogen | PEM fuel cells; solid acid fuel cells | Hydrogen infrastructure; cathodes, anodes; membranes | Hydrogen generation; catalysts | | Maintenance |
| Hydro & marine power | Hydrokinetic power systems | Wave and tide energy converters | | | |
| Geothermal | Power plants | Heat exchangers | | | Installation |
| Nuclear | Modular reactors; fusion | | | | |
| Air | Carbon capture | | Pollutant detection; gas treatment | Sensors; satellites; measuring air quality | |
| Advanced materials | | Feedstock for 3D printing; 3D printers | Fibers; nanoparticles; manufacturing; materials; membranes; measurement techniques | Software for material selection; manufacturing improvements | |
| Other cleantech | Projects; prefabricated homes | | R&D; manufacturing | Software platforms; modeling; smart technologies; | Big data analytics; security; audits; consulting |

| | |
|---|---|
| | metering; IoT;<br>power management |

**SUPPLEMENTARY NOTE 4: Changes in investments between 2006–2013 and 2014–2021**

Supplementary Table 4 shows the changes in early-stage private investments between 2006–2013 and 2014–2021 for each sector (corresponding to information in Figure 1). In the latter period, i.e., after 2012, early-stage private investments in the transportation and agriculture sectors experienced the largest increase, while the solar sector saw the largest decrease.

***Supplementary Table 4:*** Change in private investments in each sector between 2006–2013 and 2014–2021. The fourth column, Delta ($ million), is the difference between the first two numerical columns and corresponds to the data presented in the rightmost "total" plot in Figure 2.

| Sector | 2006–2013 ($ million) | 2014–2021 ($ million) | Percent change in sector | Delta ($ million) | Percent change compared to all-sector total |
|---|---|---|---|---|---|
| Energy efficiency | 1,243 | 1,028 | -17.2% | -214 | -2.2% |
| Solar | 826 | 398 | -51.8% | -428 | -4.4% |
| Wind | 125 | 48 | -62.0% | -77 | -0.8% |
| Energy storage | 303 | 392 | 29.3% | 89 | 1.0% |
| Smart grid | 235 | 229 | -2.5% | -6 | 0.0% |
| Agriculture | 423 | 2,518 | 495.1% | 2,095 | 21.4% |
| Transportation | 647 | 4,034 | 523.8% | 3,387 | 34.8% |
| Fuel cells & hydrogen | 43 | 73 | 68.3% | 29 | 0.4% |
| Hydro & marine power | 21 | 11 | -46.9% | -10 | -0.2% |
| Geothermal | 12 | 61 | 394.1% | 49 | 0.6% |
| Nuclear | 39 | 117 | 203.4% | 78 | 0.8% |
| Air | 171 | 120 | -30.1% | -51 | -0.6% |
| Advanced materials | 505 | 757 | 49.9% | 252 | 2.6% |
| Other cleantech | 603 | 5,164 | 755.9% | 4,561 | 46.8% |
| Total | 5,196 | 14,950 | 187.7% | 9,754 | 100% |

**SUPPLEMENTARY NOTE 5: Indicators of innovation by technology sector and product cluster**

Our scoping analysis (Figures 1–2) included 14 of the technology sectors in the i3 database, and the main manuscript also includes graphs and detailed analysis for three technology sectors that exemplified investment trends in different phases of the early innovation life cycle. Supplementary

Figures S3–S16 present data on all 14 technology sectors included in our analysis, i.e., solar power, wind power, energy storage, transportation, fuel cells & hydrogen, hydro & marine power, geothermal, nuclear, air, advanced materials, and other cleantech. In addition to the three sectors discussed in the main paper (energy efficiency, energy storage, and agriculture), we also describe trends in more detail for four other sectors (solar, wind, smart grids, and transportation).

We exclude grants and number of investors in the main manuscript (see Figures 3-5). Grants are indicators of the public policies that incentivize investments, so we separate these to identify the effect of these policies on early-stage private investments. We found that the number of investors was highly correlated with the number of firms (the number of investors is higher), and thus this metric did not provide meaningful new insights as an independent indicator of innovation. We present these metrics for each product cluster in this Supplementary Note.

**Energy efficiency.**

We present our analysis of early-stage private investments in energy efficiency and firms that receive them in the main manuscript. Other metrics of investments (i.e., grant funding and number of investors) also decline after 2013: cumulative grants are 15.5% lower in the 2014–2021 time period compared to 2006–2013 and the average number of annual investors is 2% lower.

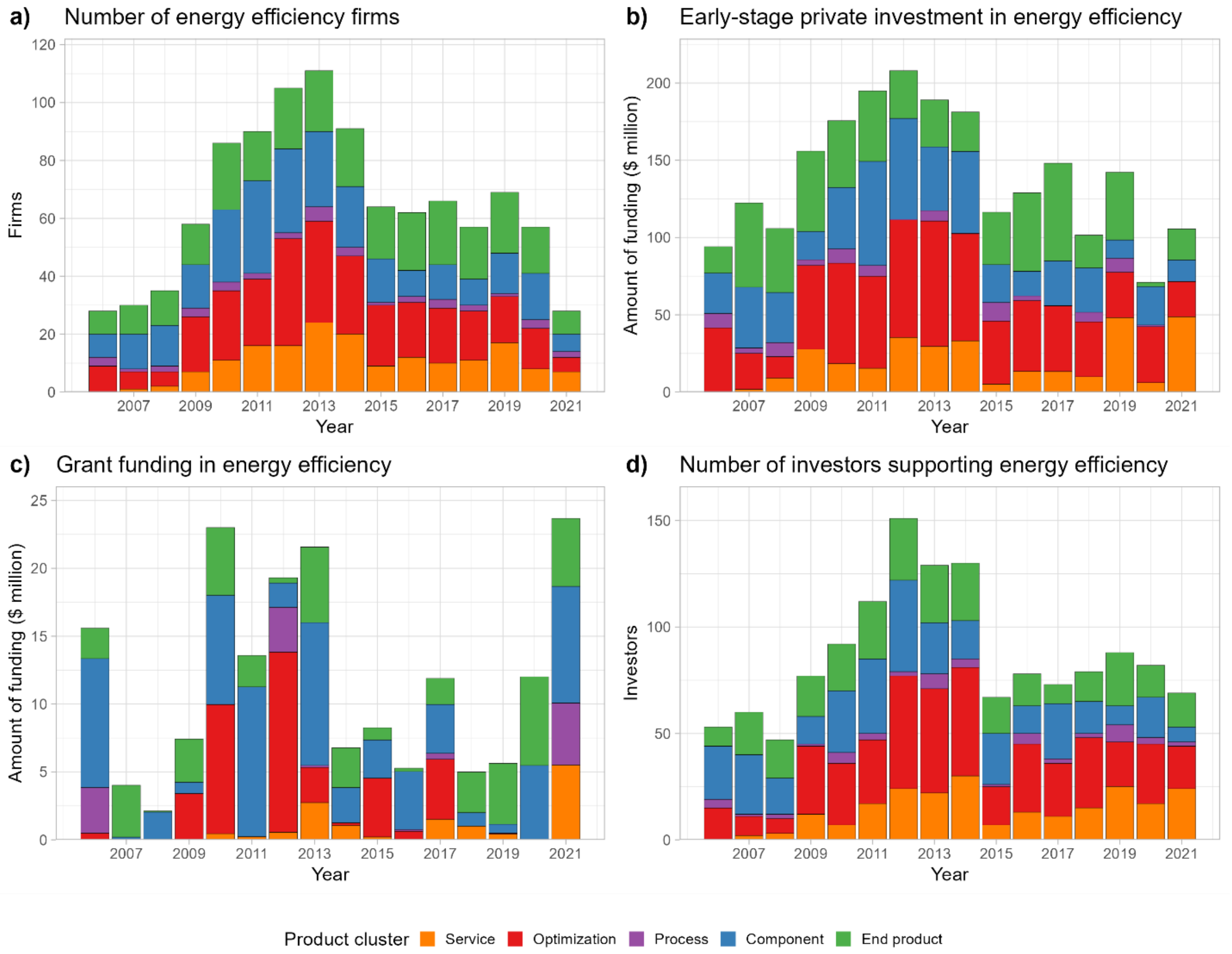

***Figure S3***: Comparison of number of firms (a), early-stage private investments (b), grants (c), and number of investors (d) between 2006 and 2021 for energy efficiency energy. Vertical axes are scaled differently in the individual charts. Colors indicate the type of product-cluster within the nascent value chain.

## Solar energy.

The pronounced decline in early-stage private sector funding for solar energy after 2011, shown in Supplementary Figure S3b, coincides with crystalline-silicon (c-Si) based photovoltaic (PV) technology becoming the dominant commercial design in the early 2010s (7,8). Our data show that the decrease in investment occurred most notably in components and processes. About 21% of the 216 firms developed products focused on alternative approaches (7), such as PV cells based on gallium arsenide (GaAs), nanostructured materials, and copper-indium gallium selenide (CIGS), or thin films, as well as concentrated solar power (CSP) (see Supplementary Figure S18). Funding for these competitors to c-Si PV dropped dramatically after 2013, when only 10% of the firms developing alternative solar technologies received early-stage private sector funding. A small increase in private sector investment is evident in business services (see Supplementary Figures S3b), primarily in new approaches to installation logistics. Government grants did not mirror the decline in private investment in components until 2015 (see Supplementary Figure S3c), with continuing support including products such as power electronics, mountings, trackers, and lighting.

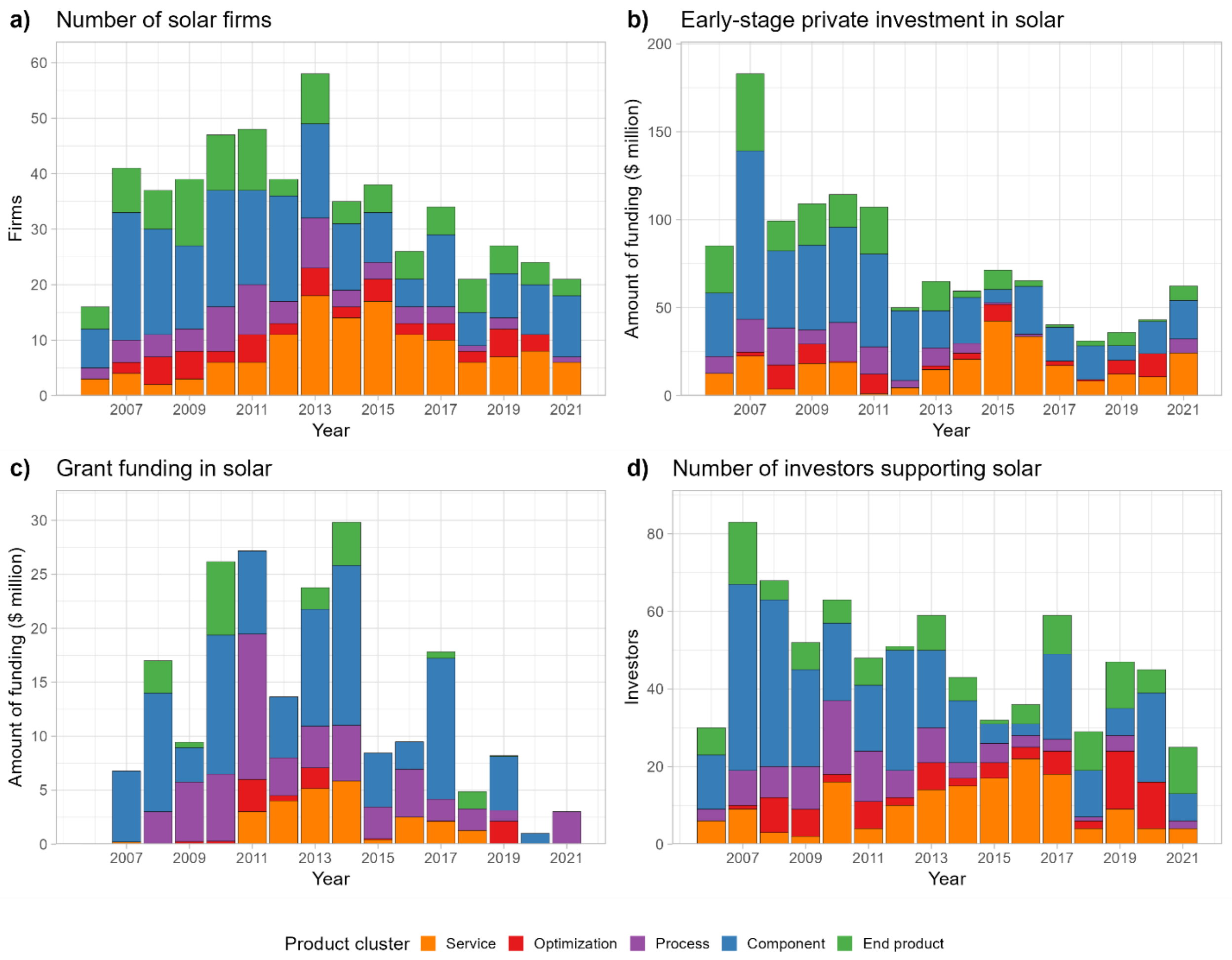

***Figure S4:*** Comparison of number of firms (a), early-stage private investments (b), grants (c), and number of investors (d) between 2006 and 2021 for solar. Vertical axes are scaled differently in the individual charts. Colors indicate the type of product-cluster within the nascent value chain.

## Wind energy.

Early-stage private investment for wind energy technologies largely focused on end products and services from 2006 to 2013, characteristic of system-component innovation (9) and then dropped in 2012 (see Supplementary Figure S5b). In onshore wind energy, three-blade, horizontal axis wind turbine design (9,10) experienced a continuing improvement in capacity factor, contributing to the decrease in the levelized cost of wind-generated power around 2011 (11). Our data show that the corresponding drop in early-stage private sector investment impacted firms focused on applications of the dominant turbine model to different conditions such as on rooftops or in wind farms, as well as those focused on alternative approaches or on more disruptive designs including very high-altitude wind extraction (see Supplementary Figure S19). Policy uncertainties related to the expiration of the production tax credit — the most impactful policy for incentivizing wind energy deployment (12) — also possibly slowed market-pull for riskier, alternative wind turbine designs.

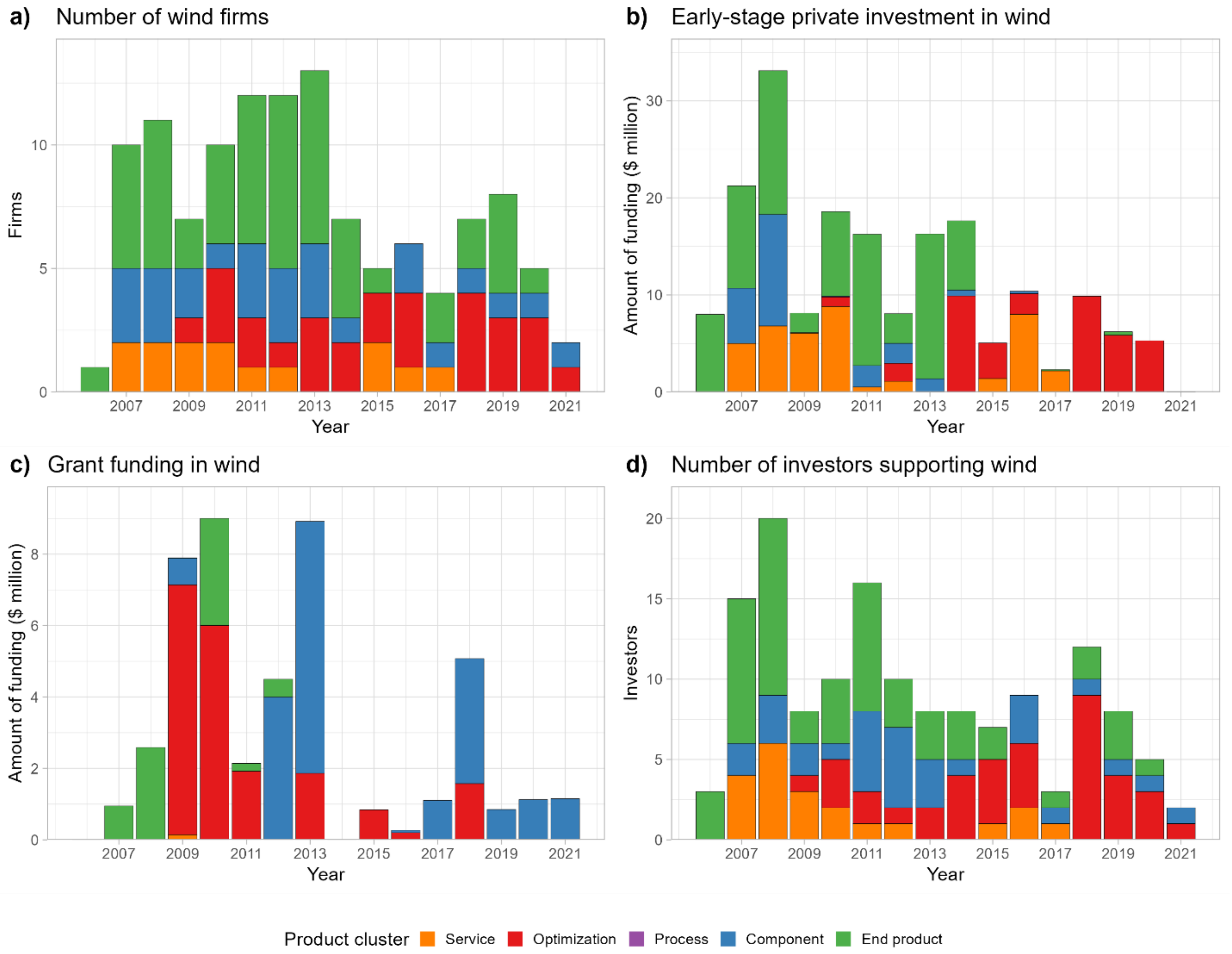

***Figure S5:*** Comparison of number of firms (a), early-stage private investments (b), grants (c), and number of investors (d) between 2006 and 2021 for wind. Vertical axes are scaled differently in the individual charts. Colors indicate the type of product-cluster within the nascent value chain.

### Energy storage.

We present our analysis of early-stage private investments in energy storage and firms that receive them in the main manuscript. Grant funding declines after 2013 but experiences a second wave 2018–2021.

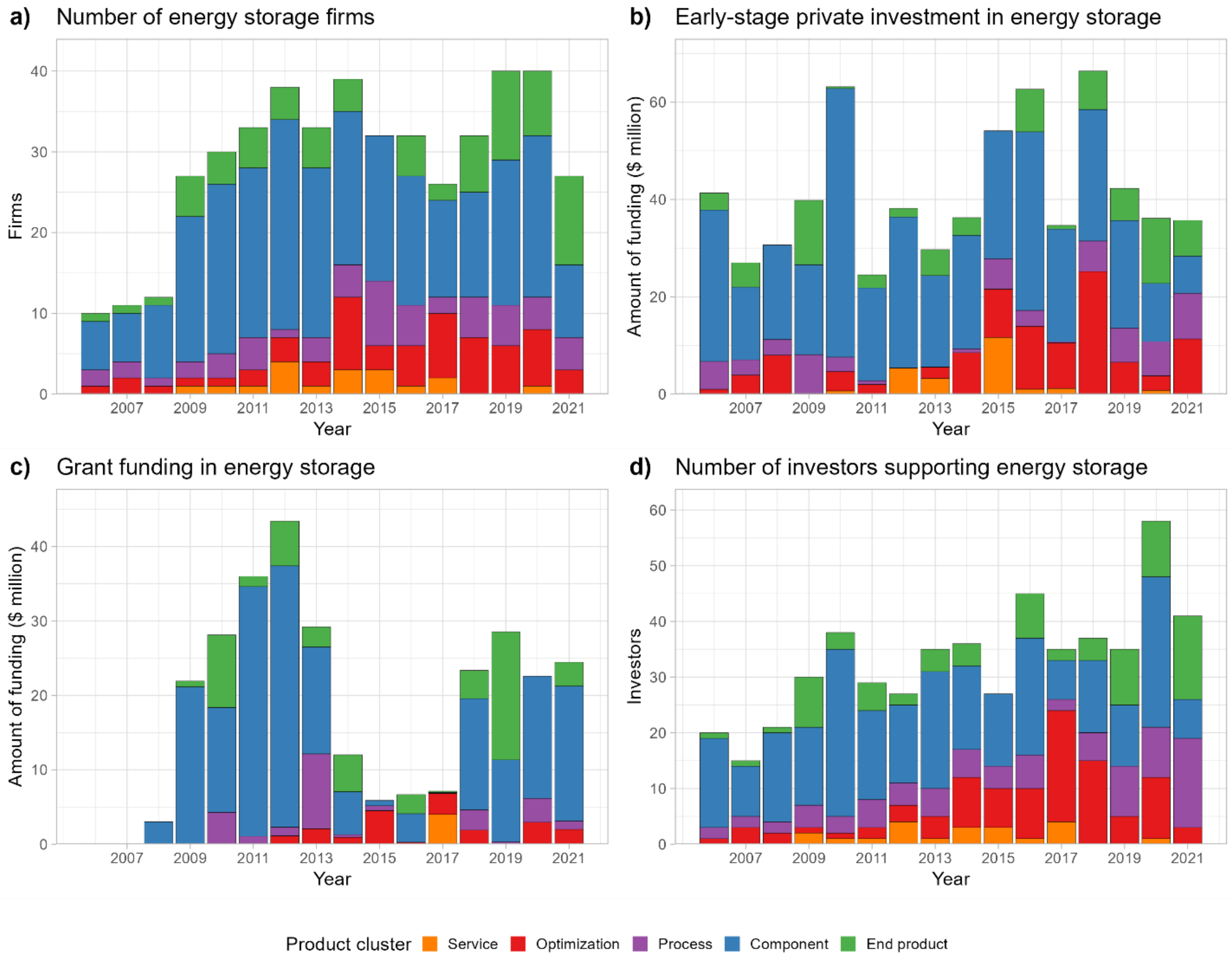

***Figure S6:*** Comparison of number of firms (a), early-stage private investments (b), grants (c), and number of investors (d) between 2006 and 2021 for energy storage. Vertical axes are scaled differently in the individual charts. Colors indicate the type of product-cluster within the nascent value chain.

## Smart grid.

The smart grid sector represents a collection of technologies and services to modernize the electric grid in both transmission and distribution (13–15). Despite interannual variability due to a smaller number of firms (113 firms compared with 496 energy efficiency firms), private sector investment in the smart grid sector was relatively consistent in our study period (see Supplementary Figure S7b), and there has been a persistent focus on optimization. Firms developing these products (see Supplementary Figure S20) take advantage of advances in computing from improvements in electricity systems modeling (16) to better capabilities for distributed (edge) computing (17). As a result of these innovations, firms prior to 2014 develop technologies to establish smart grid capabilities (e.g., developing new technologies to monitor and aggregate data about electricity consumption) while firms receiving investments after 2014 increasingly develop technologies to enhance existing smart grid capabilities (e.g., developing new methods to connect more electricity consumers to existing monitoring networks). Within the components product cluster, the development of sensors was also closely linked to optimization activities on sensor networks. This ongoing innovation is a result of continued coordination between federal and state agencies to design public policies that clearly encourage utilities to incorporate smart grid technologies in their distribution networks (18).

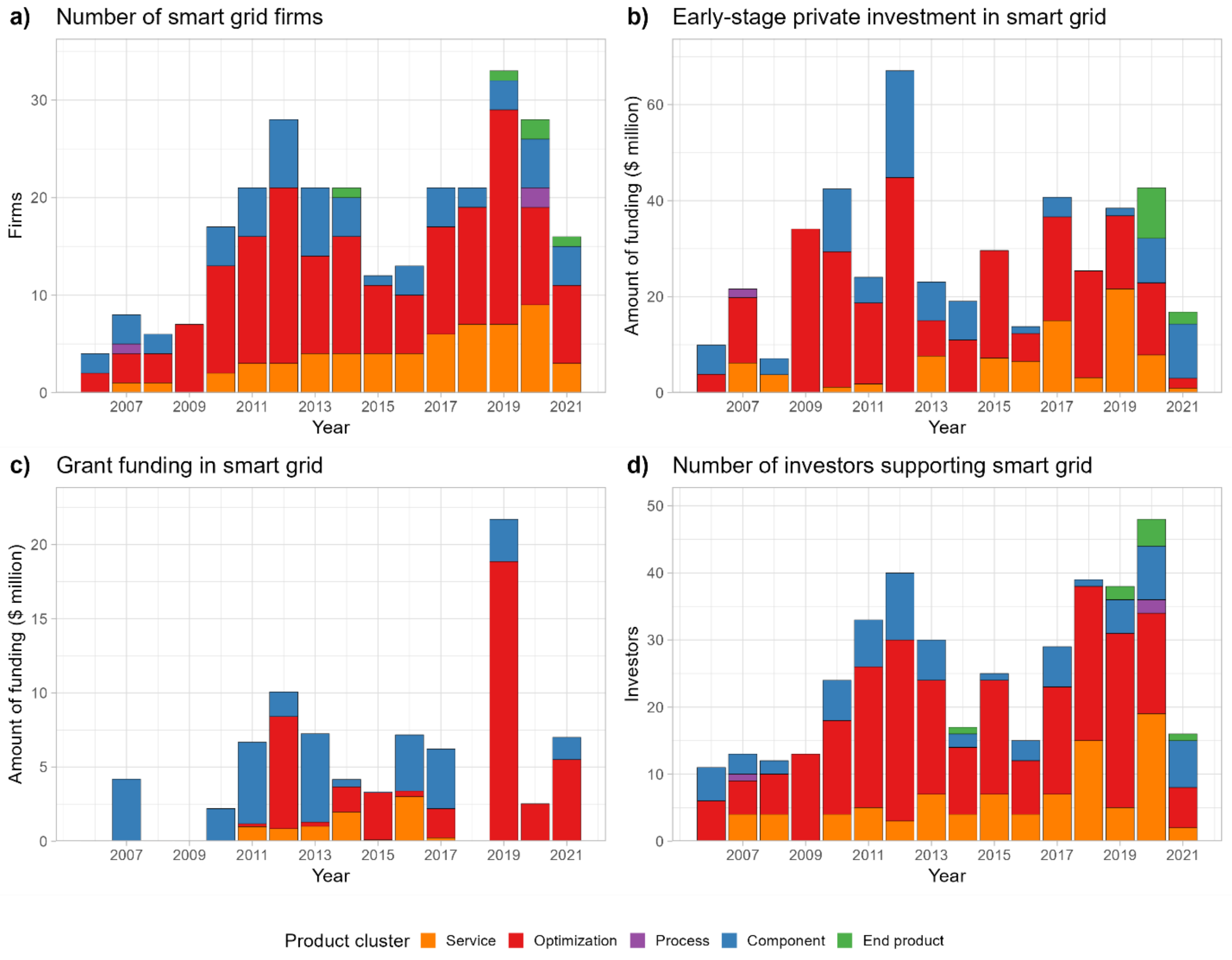

***Figure S7:*** Comparison of number of firms (a), early-stage private investments (b), grants (c), and number of investors (d) between 2006 and 2021 for smart grid. Vertical axes are scaled differently in the individual charts. Colors indicate the type of product-cluster within the nascent value chain.

## Agriculture.

We present our analysis of early-stage private investments in agriculture and firms that receive them in the main manuscript. Grant funding declines after 2013, but all other metrics show new growth after 2013.

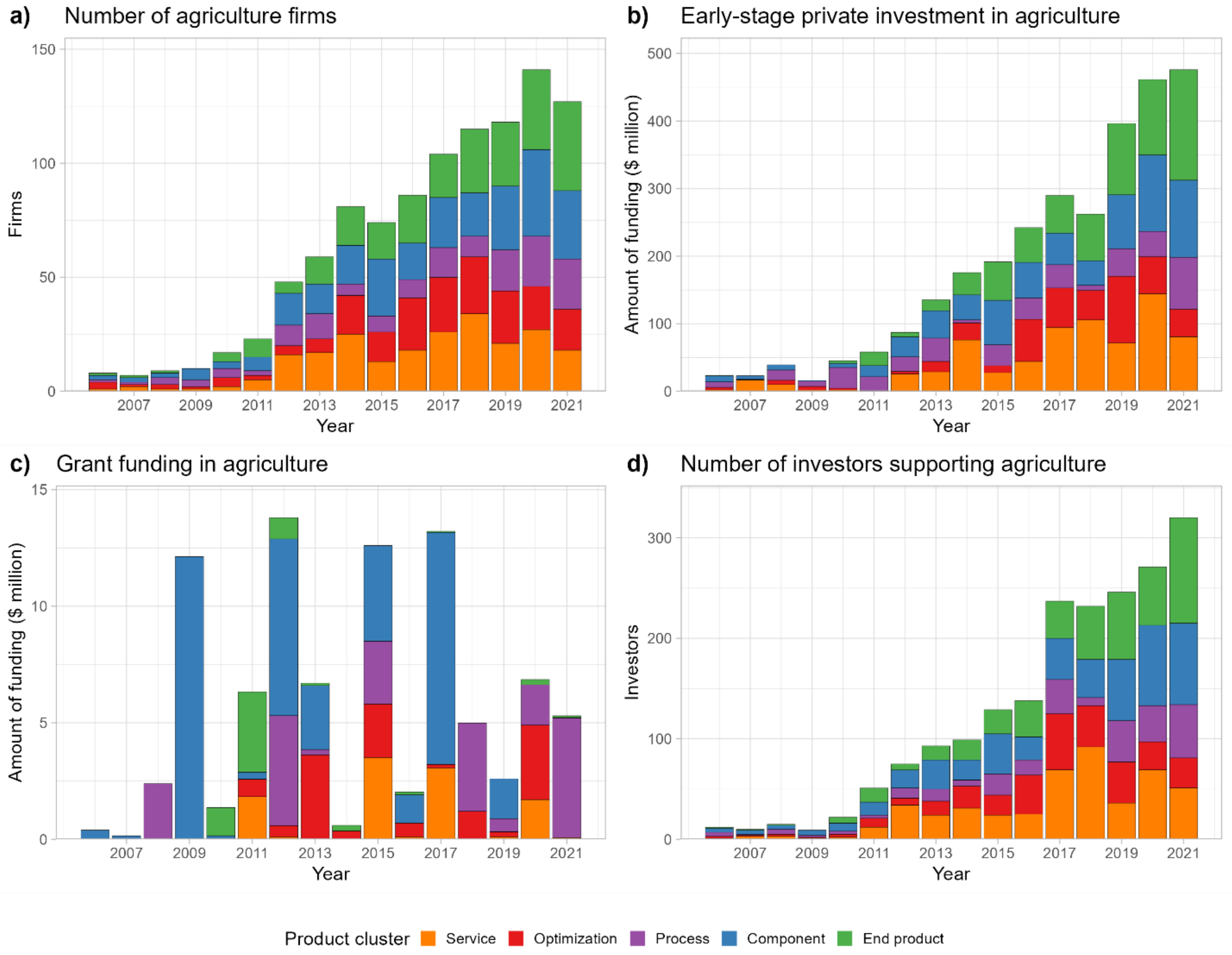

***Figure S8:*** Comparison of number of firms (a), early-stage private investments (b), grants (c), and number of investors (d) between 2006 and 2021 for agriculture. Vertical axes are scaled differently in the individual charts. Colors indicate the type of product-cluster within the nascent value chain.

## Transportation.

Early-stage private sector investments in transportation began increasing around 2011 and then grew dramatically (see Supplementary Figure S9b), increasing by a factor of 6.2 from 2006–2013 to 2014–2021. However, the individual product clusters increased by dramatically different amounts, with growth driven first by optimization (increasing by a factor of 17.4) and then services (increasing by a factor of 8.0). Over the analysis time period, 37.3% of firms are developing products in either business services (268 out of 636 firms) or optimization (90 out of 232 firms). These changes coincided with strong commercial demand for mobility services (19) and growing recognition of the technological opportunities for vehicle automation (20) (see Supplementary Figure S23). Traditionally, the climate-innovation focus for transportation has been on reducing emissions directly through improved efficiency, electrification, and alternative fuels. This is reflected in the components product cluster, where investment levels were essentially constant over the study period. However, the products in the services and optimization product clusters represent a distinct change in the nascent value chain. These mobility and automation investments are often included as climate-tech, but their emissions reduction potential remains to be demonstrated (1,20,21). There are technical opportunities for delivering emissions reduction within the new transportation trends (22,23), and some applications may help spur demand for electric vehicles (24). Public policy efforts will be required to bolster the emissions reduction benefits of the recent commercial trends (25,26).

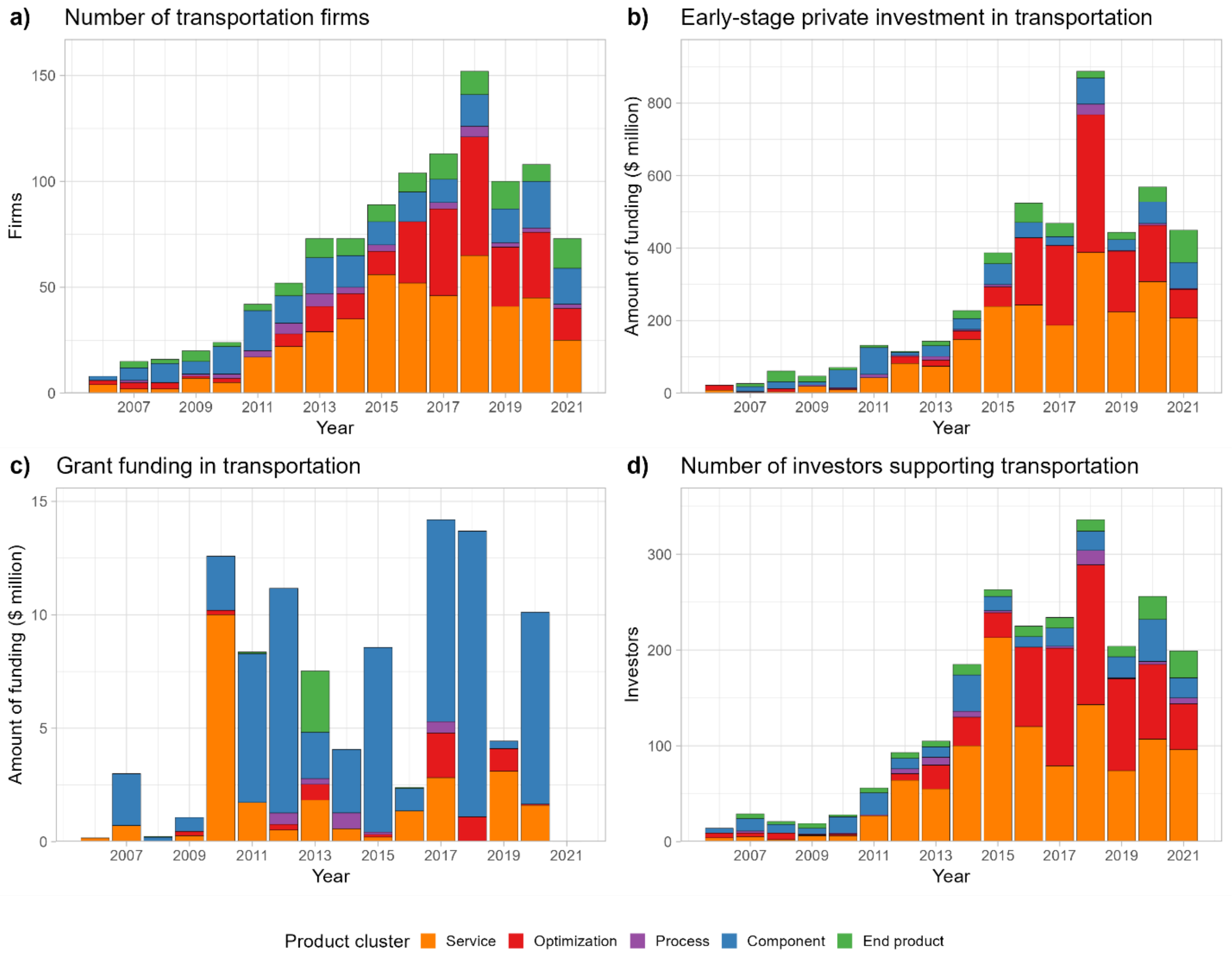

***Figure S9:*** Comparison of number of firms (a), early-stage private investments (b), grants (c), and number of investors (d) between 2006 and 2021 for transportation. Vertical axes are scaled differently in the individual charts. Colors indicate the type of product-cluster within the nascent value chain.

## Fuel cells & hydrogen.

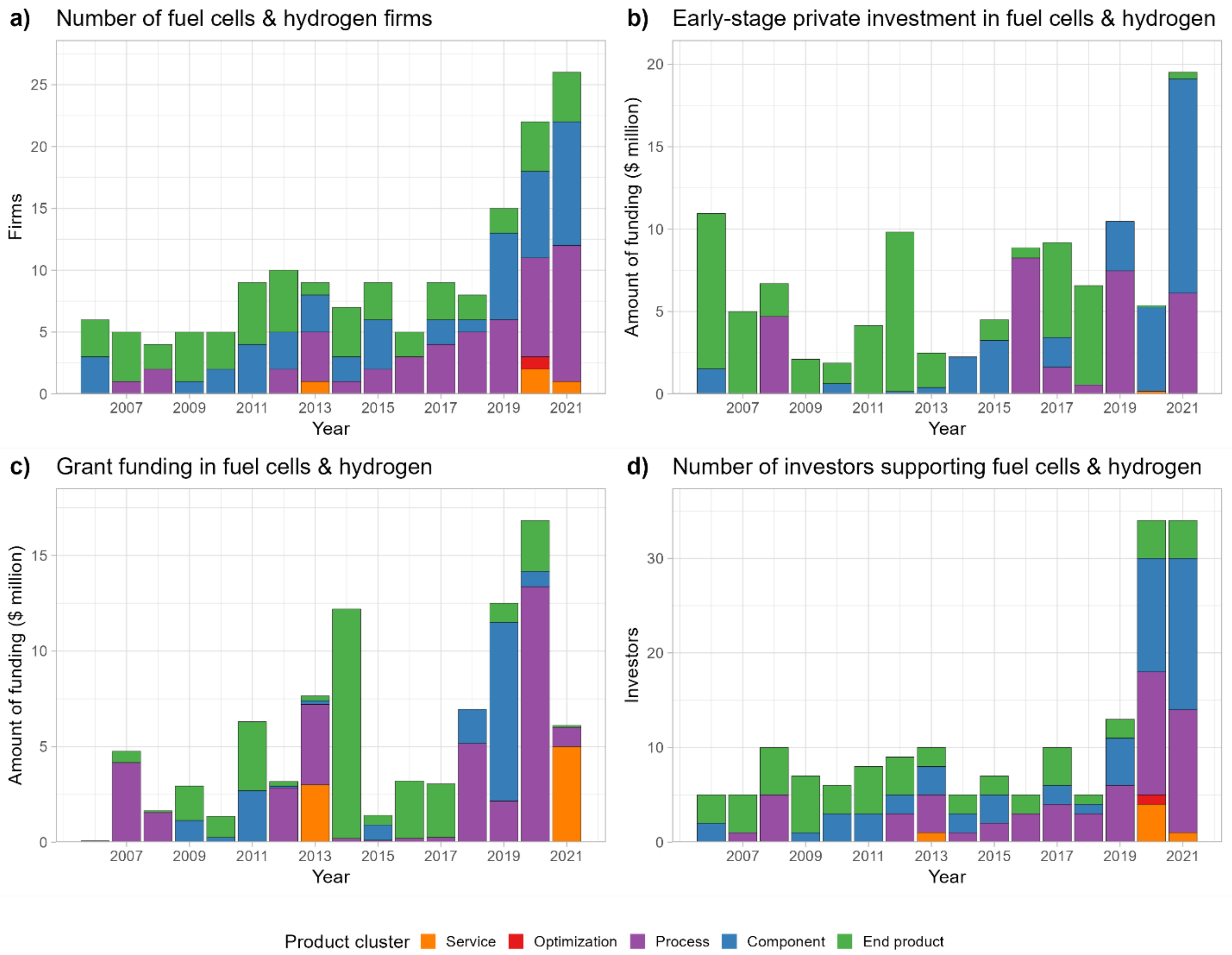

***Figure S10*:** Comparison of number of firms (a), early-stage private investments (b), grants (c), and number of investors (d) between 2006 and 2021 for fuel cells & hydrogen. Vertical axes are scaled differently in the individual charts. Colors indicate the type of product-cluster within the nascent value chain.

## Hydro & marine power.

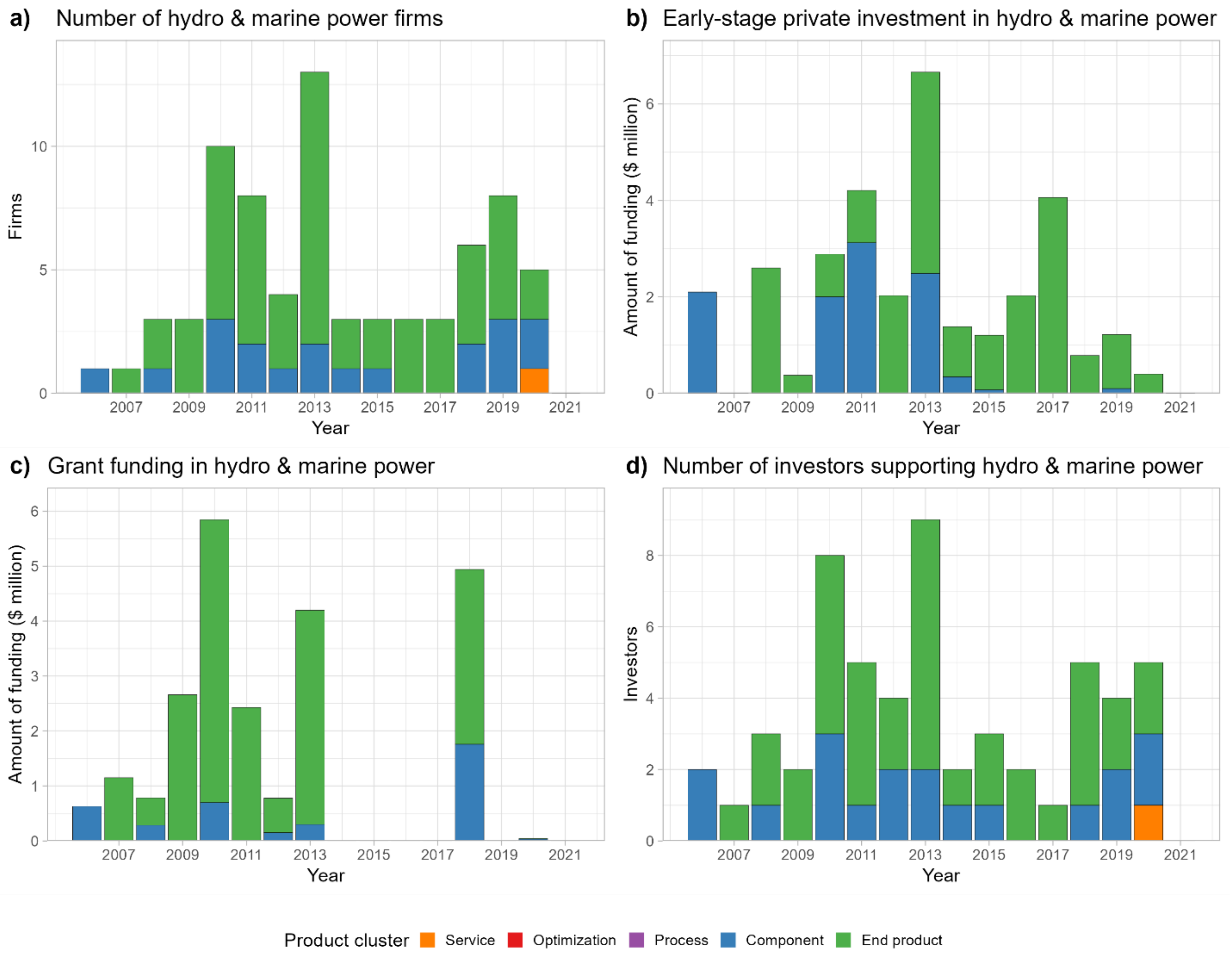

***Figure S11:*** Comparison of number of firms (a), early-stage private investments (b), grants (c), and number of investors (d) between 2006 and 2021 for hydro & marine power. Vertical axes are scaled differently in the individual charts. Colors indicate the type of product-cluster within the nascent value chain.

## Geothermal energy.

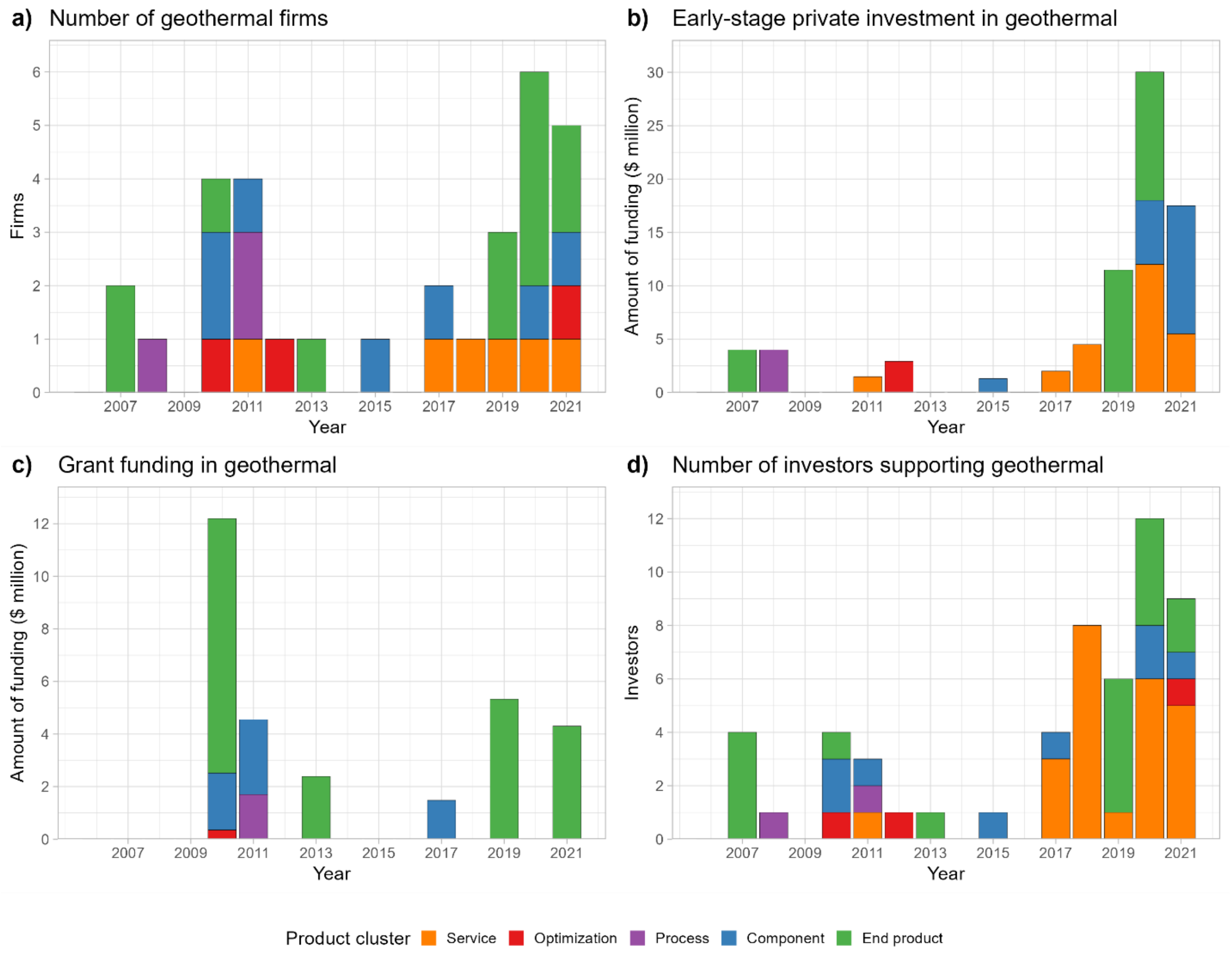

***Figure S12:*** Comparison of number of firms (a), early-stage private investments (b), grants (c), and number of investors (d) between 2006 and 2021 for geothermal energy. Vertical axes are scaled differently in the individual charts. Colors indicate the type of product-cluster within the nascent value chain.

**Nuclear.**

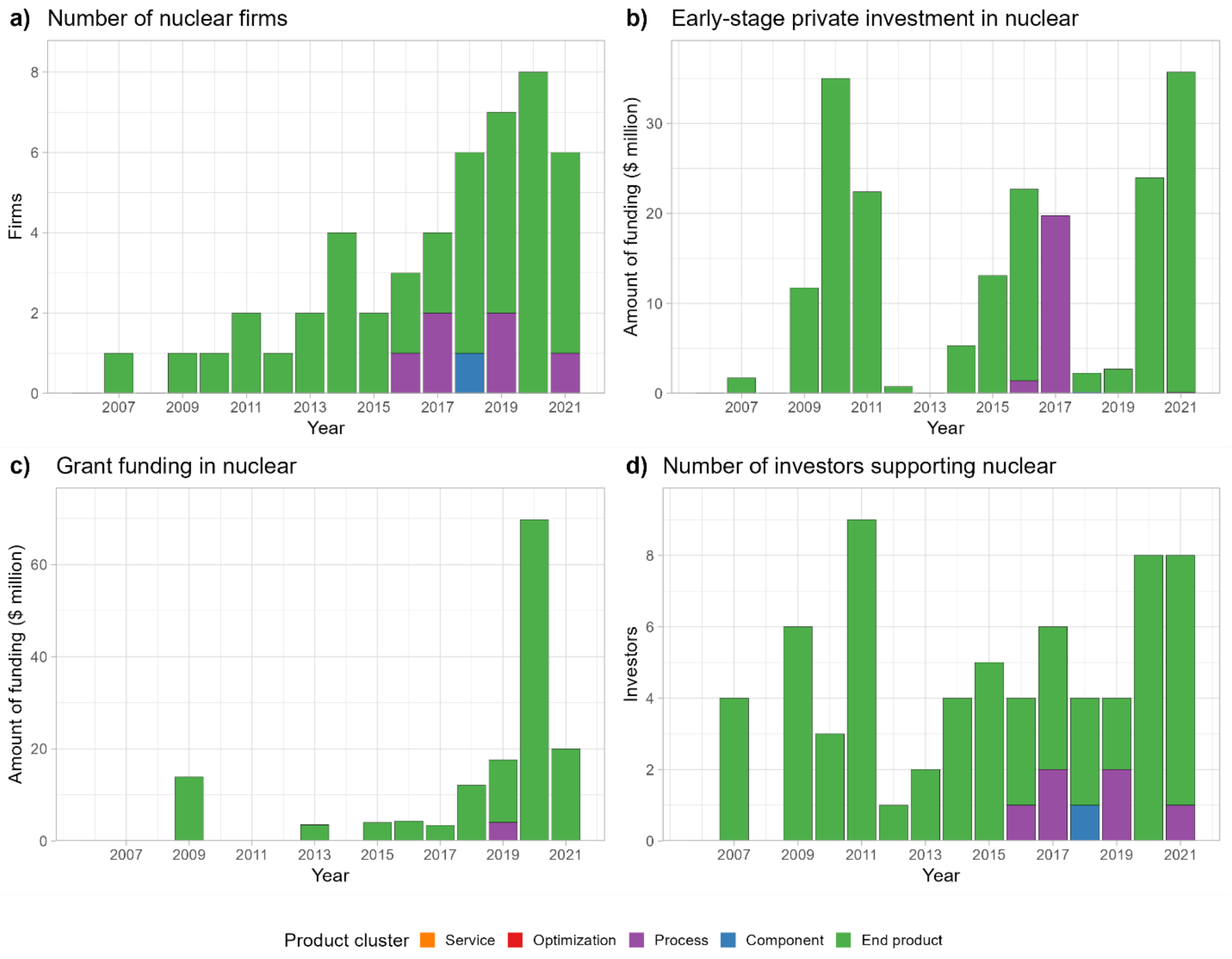

***Figure S13:*** Comparison of number of firms (a), early-stage private investments (b), grants (c), and number of investors (d) between 2006 and 2021 for nuclear energy. Vertical axes are scaled differently in the individual charts. Colors indicate the type of product-cluster within the nascent value chain.

**Air.**

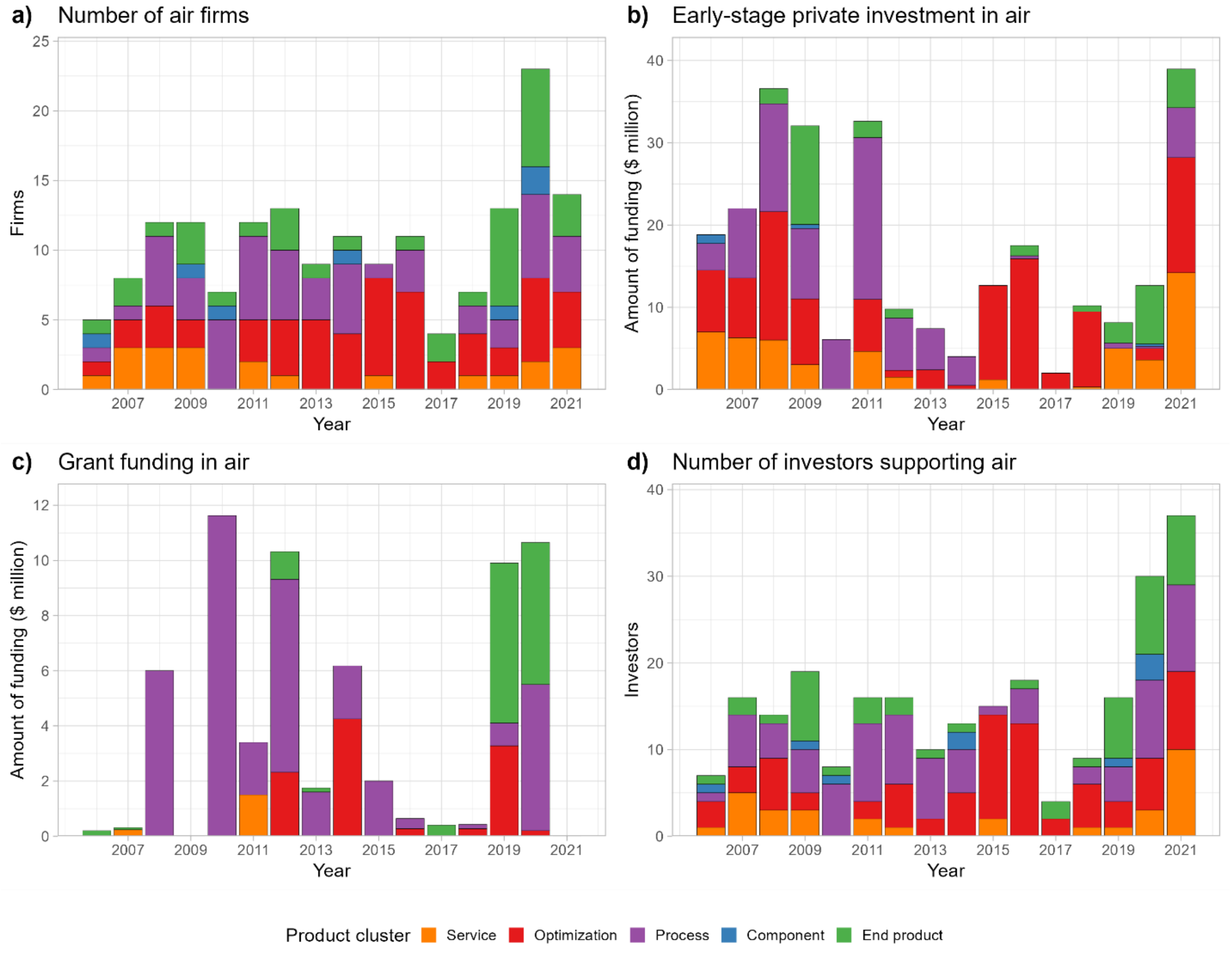

***Figure S14:*** Comparison of number of firms (a), early-stage private investments (b), grants (c), and number of investors (d) between 2006 and 2021 for air. Vertical axes are scaled differently in the individual charts. Colors indicate the type of product-cluster within the nascent value chain.

## Advanced materials.

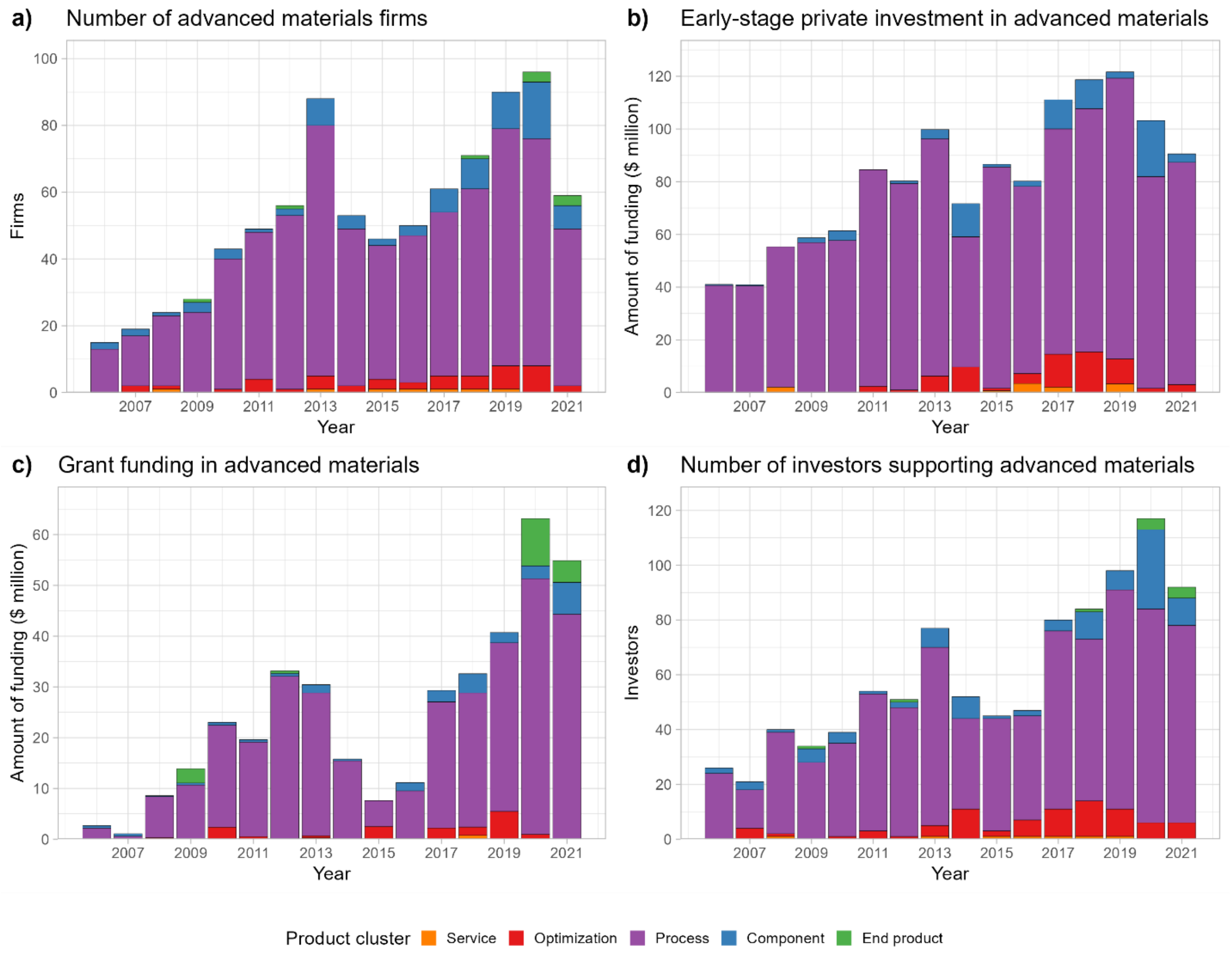

***Figure S15:*** Comparison of number of firms (a), early-stage private investments (b), grants (c), and number of investors (d) between 2006 and 2021 for advanced materials. Vertical axes are scaled differently in the individual charts. Colors indicate the type of product-cluster within the nascent value chain.

**Other cleantech.**

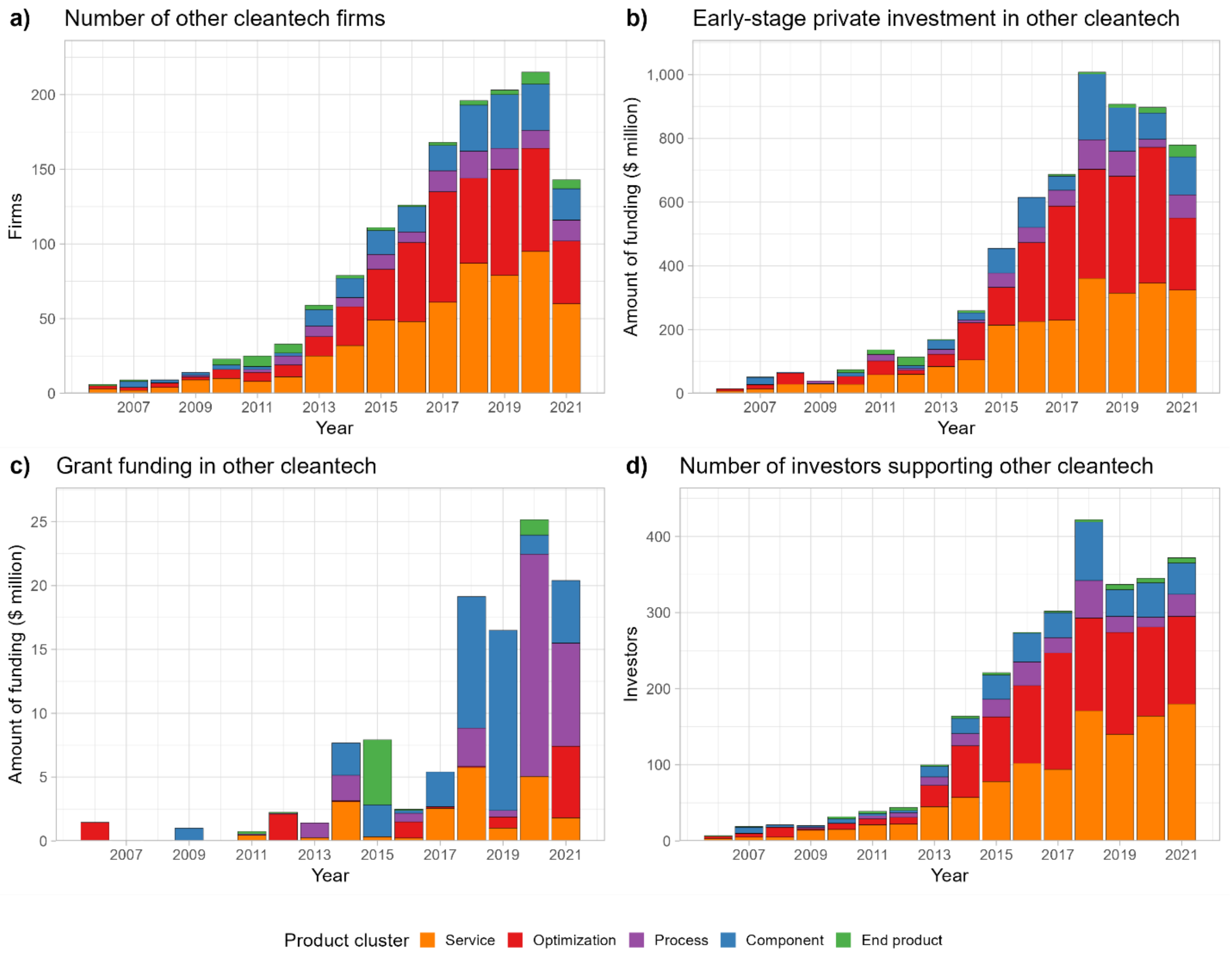

***Figure S16:*** Comparison of number of firms (a), early-stage private investments (b), grants (c), and number of investors (d) between 2006 and 2021 for other cleantech. Vertical axes are scaled differently in the individual charts. Colors indicate the type of product-cluster within the nascent value chain.

## SUPPLEMENTARY NOTE 6: Product differentiation in selected technology sectors

The manuscript discusses the context in which selected examples from the technology sectors developed and includes information on specific products within product clusters most relevant to the underlying trends in the technology sector. Supplementary Figures S17–S23 present additional results for some of the specific products for each of the seven sectors that we analyze in more detail, including the dollar amount of early-stage private investments received by firms and the number of firms that received such investment during the 2006–2021 time period. For each sector, this Supplementary Note describes how investments in the relevant products evolved over the time period.

## Energy efficiency.

Firms in the energy efficiency sector have received declining early-stage private investments from 2006–2021. The largest group of recipients of early-stage private investments in the energy efficiency sector develop building efficiency products. These are shown in Supplementary Figure S17, which represents 331 firms, receiving $1,599 million, out of all 496 energy efficiency firms receiving $2,271 million. The four largest applications of energy efficiency technology in buildings are internet-connected appliances and "smart" technology broadly called the *Internet-of-Things* (IoT), *lighting* products such as LEDs and lighting systems, *climate* products such as HVAC systems and refrigeration, and *other* building efficiency products such as building retrofits and construction improvements. A large fraction of the products in these areas are in the optimization product cluster and deliver Energy Management and Control Systems (EMCS) products that are based on software for building efficiency (see Figure 3). There are 120 EMCS-optimization firms receiving $560 million in early-stage private investments (shown in Supplementary Figure S17), out of 165 optimization firms in the energy efficiency sector overall.

The funding levels for the 331 firms shown in Supplementary Figure S17 have declined. Prior to 2014, 182 of these firms received $870 million in early-stage private investments, which declined to 186 firms receiving $729 million. Products related to lighting, climate, waste heat, and data storage experienced declines in early-stage private investment between 20% and 73% after 2014 relative to before. Funding for products related to IoT increased 33%.

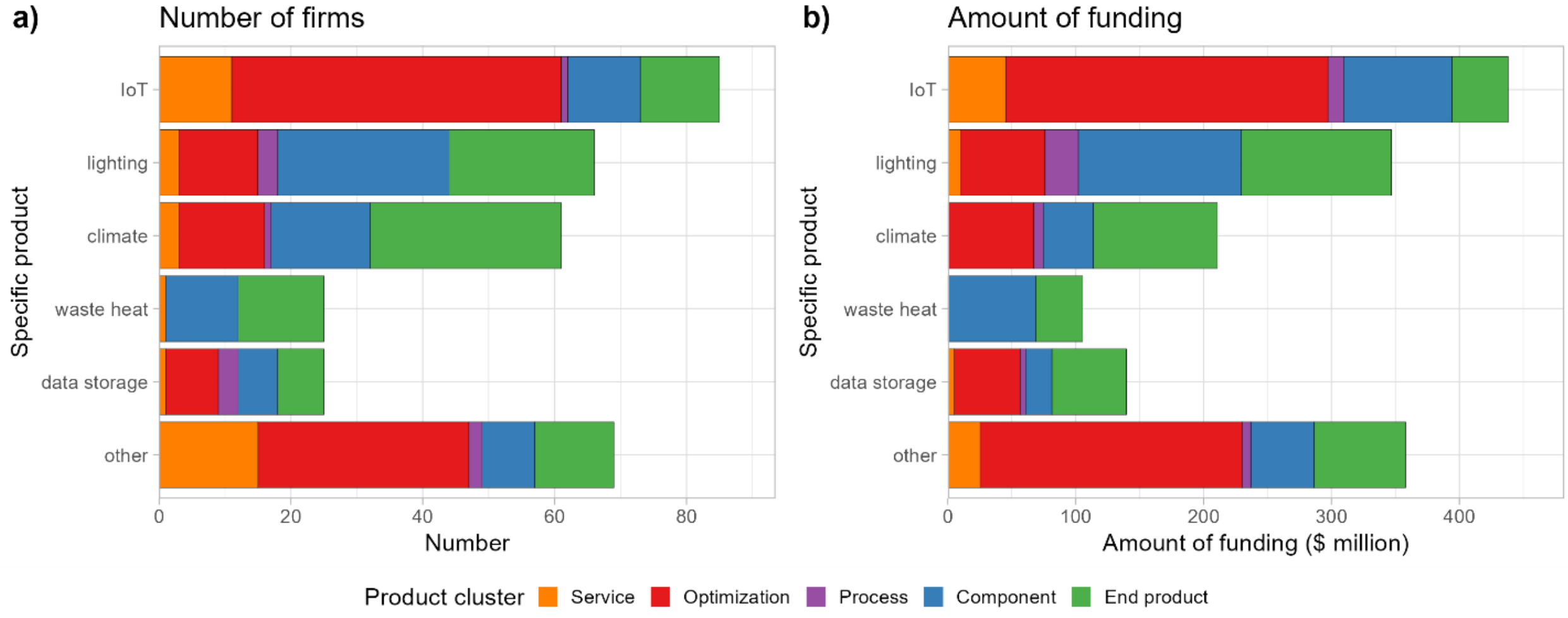

***Figure S17***: Comparison of early-stage private investments and grants for selected specific products between 2006 and 2021 by product cluster for 331 energy efficiency firms developing building efficiency products.

**Solar.**

The decrease in early-stage private investments for solar energy corresponds with crystalline-silicon (c-Si) based photovoltaic (PV) technology becoming established as the dominant commercial design in the early 2010s. Our analysis of specific products within the solar sector emphasizes this observation—Supplementary Figure S18 illustrates the distribution of 59 firms developing crystalline-silicon (c-Si) cells, and alternative solar technologies—i.e., gallium arsenide (GaAs) cells, copper indium gallium selenide (CIGS) cells, nanostructured cells (nano), thin film cells, and concentrated solar power (CSP)(27) technology. These 59 firms (out of the 269 firms in the solar sector) received $364 million in early-stage private investments between 2006–2021 out of the $1,224 million total for the solar sector. Of this, 83% of the early-stage private investments were received before 2014, including investments in 4 out of 7 firms developing c-Si solar cells, 20 out 20 of the firms developing CSP, and 28 of the 32 firms operating in the remaining cell technologies.

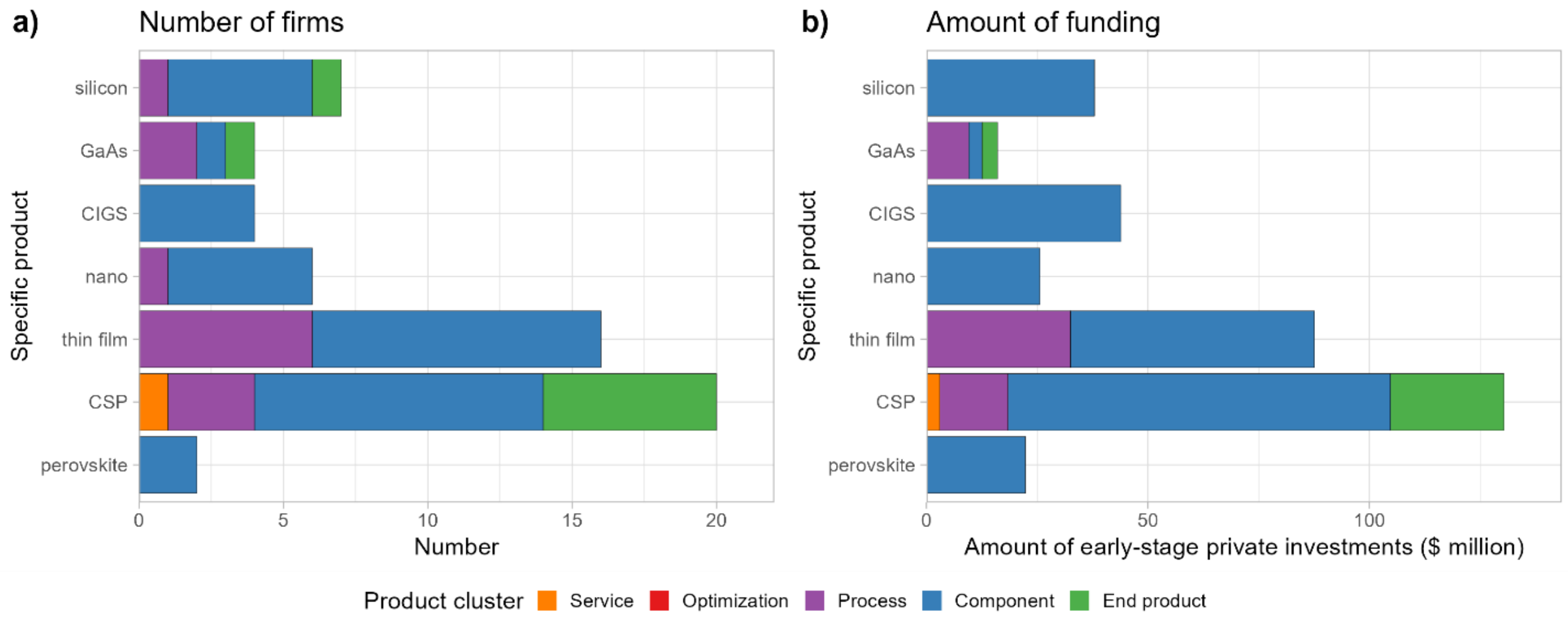

***Figure S18:*** Comparison of early-stage private investments and firms for selected specific products between 2006 and 2021 by product cluster for 59 firms developing alternative solar cells.

**Wind.**

The three-blade, horizontal axis turbine design has been dominant since the mid-1980s. Supplementary Figure S19 illustrates the 32 firms, which received $92 million in early-stage private investments for specific wind turbine designs and applications, as a subset of the total 61 wind firms which received $173 million in total. Of these 32 firms, 9 developed end products or components for the three-blade wind turbine design and another 4 worked on wind farm development using the three-blade design. Firms within this group of 13 received $64 million in funding before 2014 and $0.9 million in or after 2014. Fewer firms focused on alternative designs such as high-altitude wind extraction (2 firms), vertical axis (5 firms), or other, unique turbine designs (10 firms). Firms within this group of 17 alternative design firms, received $170 million in funding before 2014, and $192 million in funding in or after 2014.

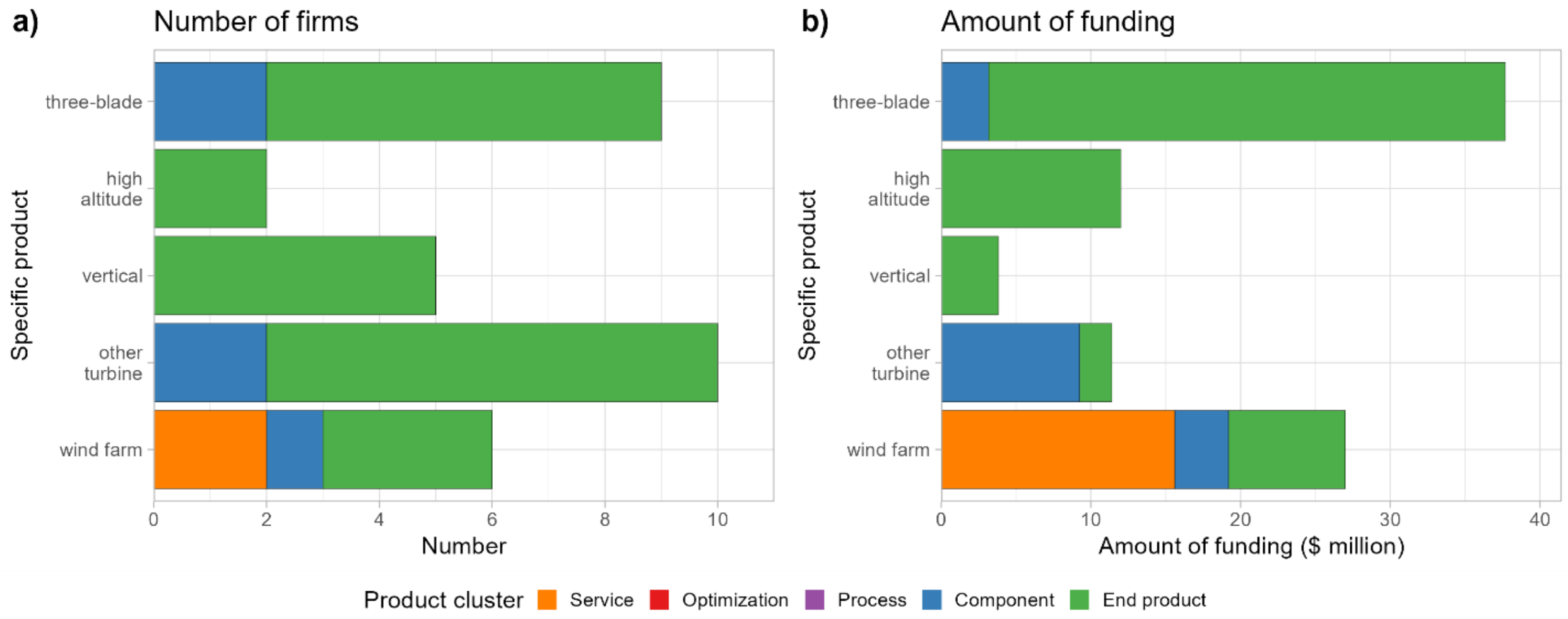

***Figure S19:*** Comparison of early-stage private investments and firms for selected specific products between 2006 and 2021 by product cluster for 61 firms developing alternative wind turbine designs and wind farms.

**Energy storage.**

Early-stage private investment temporarily decreased after 2012, as shown in Supplementary Figure S7b, coinciding with decreases in price for Li-ion batteries. Supplementary Figure S21 presents information for 106 firms (out of the total 189 firms receiving early-stage private investments) working on either Li-ion batteries, or with products that are competitive with Li-ion batteries. These firms received $433 million early-stage private investments out of the total $695 million for all the firms in the database. Of this amount, the majority was received before 2014 ($227 million). Half of these firms (53 firms) worked on Li-ion storage while the other half worked on other battery types (53 firms) (i.e., electrochemical batteries excluding lithium ion, such as lead acid, nickel hydrogen, or zinc based, as well as flow-cell batteries). 35 of these 53 lithium-ion firms received early-stage private investments after 2014 (increasing 8% to 103 million), while 30 firms working on other battery chemistries received funding (decreasing 22% to 103 million). Among firms developing Li-ion batteries, firms developing end products received more funding after 2014 ($13 million compared to $0 before), firms developing components received 52% less (at $37 million), and firms developing processes received 190% more (at $50 million). The remaining four products in Supplementary Figure S20 are energy storage technologies other than electrochemical batteries or flow cells: capacitors, thermal storage,

compressed air storage, flywheels, and pumped hydro. Funding for these firms also decreased after 2014.

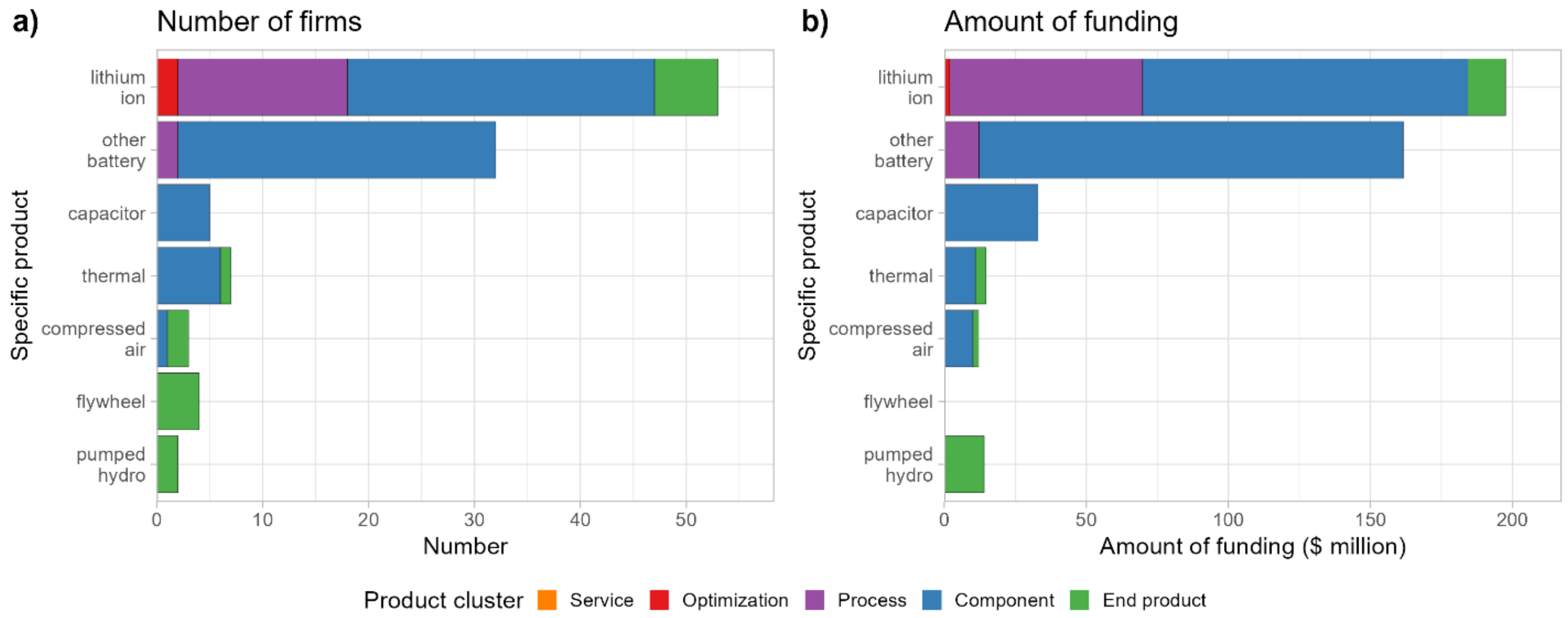

***Figure S20:*** Comparison of early-stage private investments and grants for selected specific products between 2006 and 2021 by product cluster for 106 firms developing alternative energy storage products.

**Smart Grid.**

Firms working on smart grid products focused on goals such as linking evolving population centers with generation, accommodating distributed energy resources and demand response, and maintaining reliability and security. To illustrate these goals, we identified specific products based on how they contribute to energy delivery: data collection such as sensing, measurement, and communication; demand response such management and control of electricity consumption, either supply-side or demand-side; distributed energy resources (DER) such as distributed generation, microgrids, and storage integration; and delivery of power such as distribution, transmission, and system security. 115 of the 133 smart grid firms produced one of these types of products, as shown in Supplementary Figure S21. These 115 firms received $371 million of the total $464 million early-stage private investments in the smart grid sector. Roughly half (48%) early-stage private investments in firms developing products for data collection, demand-response, and distributed energy resources were received prior to 2014.

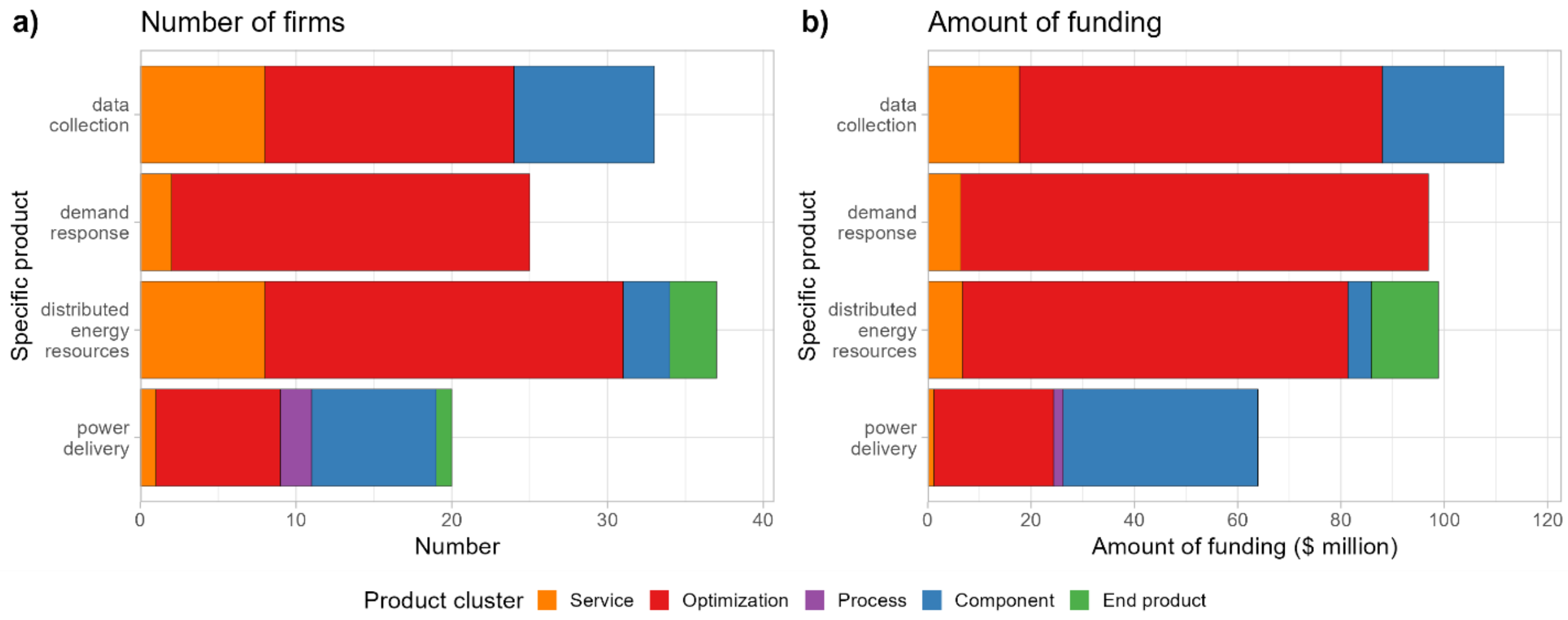

***Figure S21:*** Comparison of early-stage private investments and grants for selected specific products between 2006 and 2021 by products and product clusters for 115 smart grid firms.

**Agriculture.**

The large recent increase in funding in the agriculture sector is more balanced among the product clusters than is the case for the transportation sector. To evaluate the underpinning technical trends, we identified six specific product types that occurred frequently in the database. *Farm logistics* includes farm management, precision agriculture, and field robotics. *Delivery logistics* include preservatives and wholesale product transport and delivery services. *Livestock* includes animal feed, vaccines, and management. *Aquaculture* includes fish farming and hydroponics. *Crop treatment* includes fertilizer, pesticides, herbicides, and soil treatments. *Alternative horticulture* includes arrangements of growing crops other than traditional farms like hydroponics and vertical agriculture. *Alternative food* includes plant-based proteins or insect-based flour. *Genomics* include seed treatments and selective breeding or gene editing in plants or animals. Supplementary Figure S22 illustrates the 535 agriculture firms (out of a total of 595) producing these six types of products. These companies received $2,607 million in early-stage private investments out of the total $2,941 million for the agricultural sector overall.

The first two specific types of products, wholesale delivery logistics and farm logistics, saw an increase in technical capabilities derived from machine learning, artificial intelligence and robotics, similar to the transportation sector, and include 55 of the 99 optimization firms in the agriculture sector. Among the 213 firms with delivery and farm logistics products, the overall funding received was $972 million in early-stage private investments from 2014–2021, and only $120 million before 2014. The last seven specific product types include 118 of the 136 firms that develop components (defined for the agricultural sector as operational, expendable products such as seeds, fertilizers, and other field treatments). Firms developing these seven product types across all product clusters received $365 million before 2014 and $2,241 million from 2014 to 2021 across all product clusters.

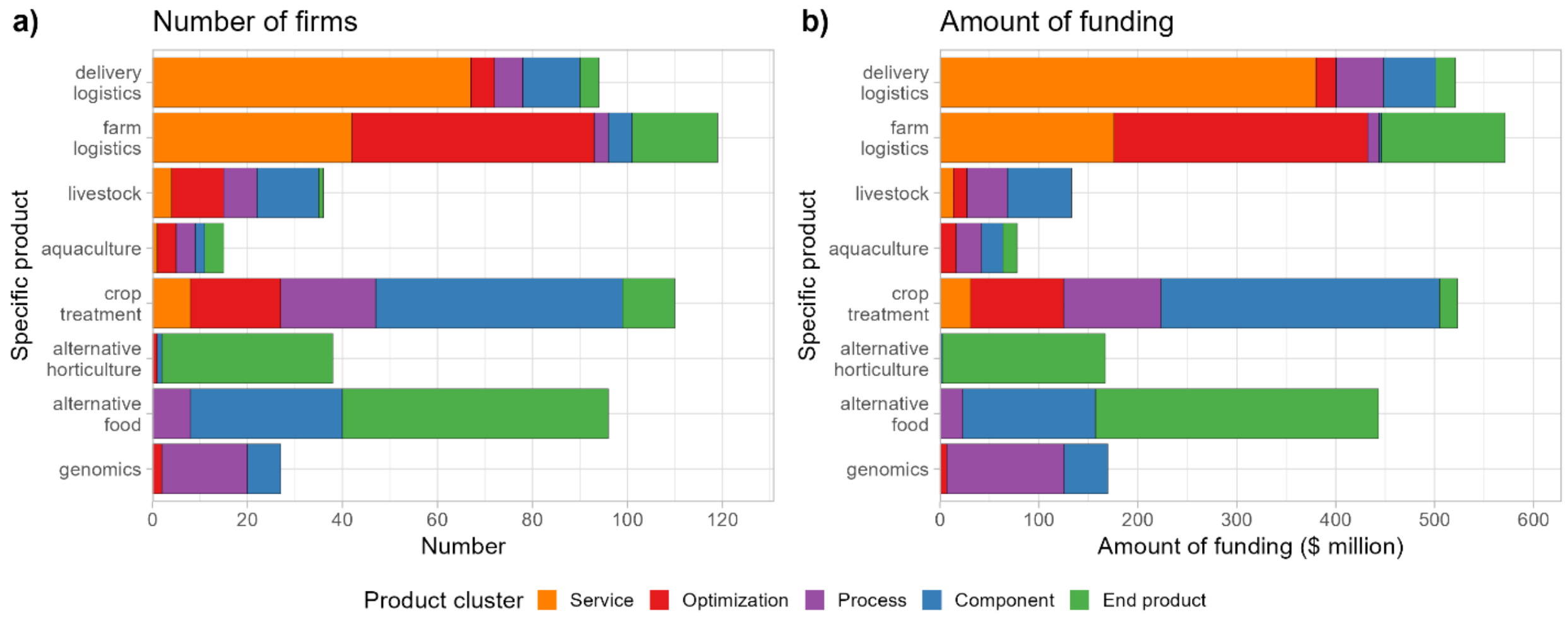

***Figure S22:*** Comparison of early-stage private investments and grants between 2006 and 2021 by product type and product cluster for 535 agriculture firms.

**Transportation.**

Most of the recent large increase in private investment in transportation occurred in the business services and optimization clusters. To evaluate this trend, we identified six specific product types: *logistics,* which include delivery, fleet management, vehicle tracking, mapping, and parking; *shared mobility* includes ride sharing, rentals, and carpooling; *autonomous vehicles* include LiDAR products, sensor, and entire vehicles, not used specifically for logistics; *fuel efficiency* includes powertrain improvements and reduced fuel consumption; *electric vehicles* (EV) include electric vehicles and charging stations; *alternative fuels* include natural gas, biofuel, and hydrogen vehicles. As shown in Supplementary Figure S23, these 6 product types represent 547 of all 602 transportation firms and $4,088 million in early-stage private investments of the total $4,680 million for the transportation sector.

Of these, the 158 firms in the last three categories, fuel efficiency, EVs, and alternative fuels were involved in developing products traditionally related to clean transportation. Funding for these product types increased from 2006–2021, with $288 million before 2014 and $515 million from 2014–2021. The first three product types in Supplementary Figure S14 (logistics, shared mobility, and autonomous vehicles), all have large product clusters in business services and/or optimization, and the 368 firms together received $291 million in early-stage private investments from 2014 to 2021, but only $2,834 million in investment before 2014.

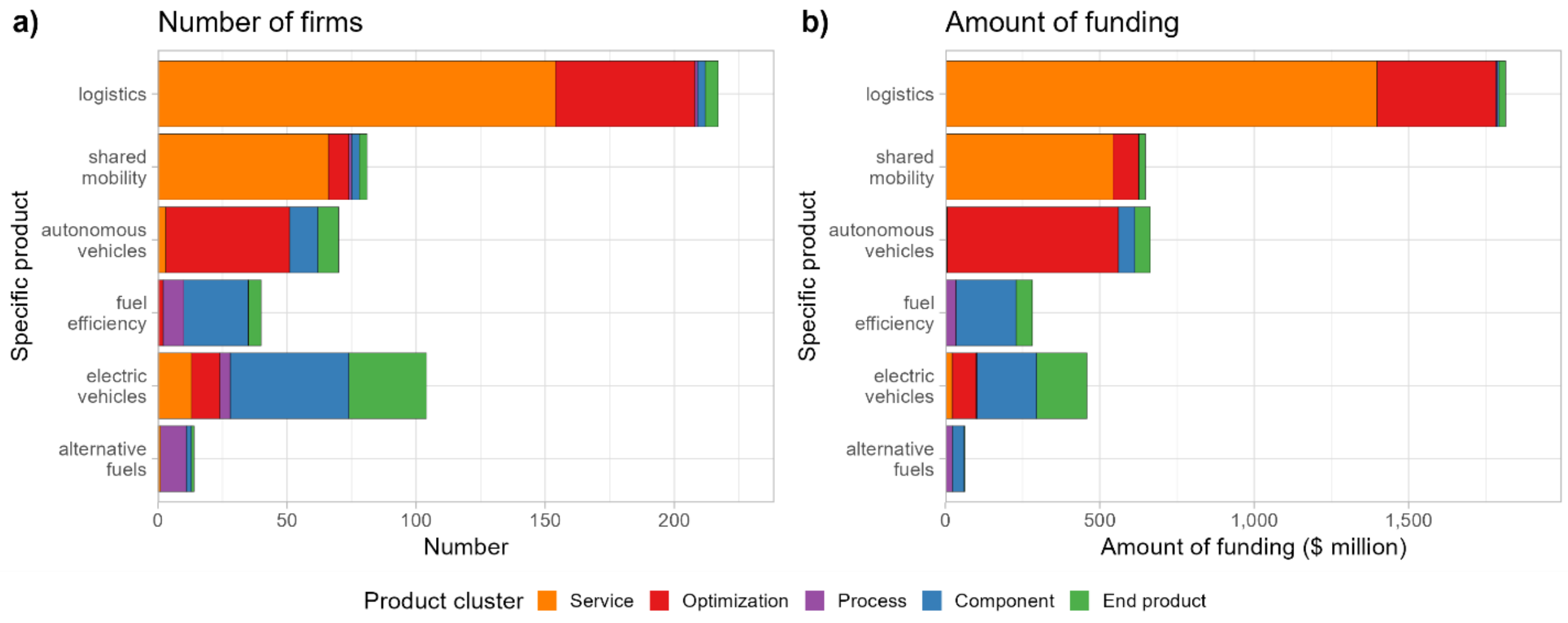

***Figure S23:*** Comparison of early-stage private investments and grants for selected specific products between 2006 and 2021 by product cluster for 547 transportation firms.

**Supplementary References**